\documentclass[11pt]{article}
\usepackage{amssymb,amsmath,amscd,amsthm,amsfonts}
\usepackage{xcolor}
\usepackage{enumerate}
\usepackage[utf8]{inputenc}
\usepackage{framed}
\usepackage{cite}
\usepackage{setspace}
\usepackage{hyperref}
\numberwithin{equation}{section}

\usepackage{fullpage}

\newcommand{\deltaM}{\delta_{\scriptscriptstyle{\rm{M}}}}

\newcommand{\half}{{{\textstyle\frac{1}{2}}}}

\newcommand{\be}{\begin{equation}}
\newcommand{\ee}{\end{equation} }
\newcommand{\beqa}{\begin{eqnarray} }
\newcommand{\eeqa}{\end{eqnarray} }
\newcommand{\ba}{\begin{array}}
\newcommand{\ea}{\end{array}}
\newcommand{\bpm}{\begin{pmatrix}}
\newcommand{\epm}{\end{pmatrix}}

\newcommand{\so}{\mathbf{so}}

\newcommand{\Spin}{\mathbf{Spin}}
\newcommand{\GL}{\mathbf{GL}}

\newcommand{\Pin}{\mathbf{Pin}}

\newcommand{\rmD}{{\rm D}}

\newcommand{\ODD}{\mathbf{O}(D,D)}

\newcommand\rd{{\rm d}}
\newcommand\rD{{\rm D}}

\newcommand\cA{{\cal A}}

\newcommand\cH{{\cal H}}

\newcommand\cJ{{\cal J}}

\newcommand\cL{{\cal L}}

\newcommand\hcL{{\hat{\cal L}}}

\newcommand\vl{{\vec{l}}}
\newcommand\vm{{\vec{m}}}

\newcommand\vbrl{{\vec{\brl}}}
\newcommand\vbrm{{\vec{\brm}}}

\newcommand\dis{\displaystyle}

\def\tx{\tilde{x}}

\def\tpartial{\tilde{\partial}}

\def\tvarphi{\tilde{\varphi}}

\def\hati{\hat{\imath}}
\def\hatj{\hat{\jmath}}
\def\hatbri{\hat{\bar{\imath}}}
\def\hatbrj{\hat{\bar{\jmath}}}
\def\hatk{\hat{k}}
\def\hatbrk{\hat{\brk}}

\def\bra{\bar{a}}
\def\brb{\bar{b}}

\def\breta{\bar{\eta}}
\def\bralpha{\bar{\alpha}}
\def\brbeta{\bar{\beta}}
\def\brgamma{\bar{\gamma}}

\def\brepsilon{\bar{\epsilon}}
\def\brrho{\bar{\rho}}

\def\brchi{\bar{\chi}}
\def\brxi{\bar{\xi}}
\def\brzeta{\bar{\zeta}}
\def\brlambda{\bar{\lambda}}
\def\brh{\bar{h}}
\def\bri{\bar{\imath}}
\def\brj{\bar{\jmath}}
\def\brk{{\bar{k}}}
\def\brl{{\bar{l}}}
\def\brm{{\bar{m}}}
\def\brn{{\bar{n}}}
\def\brp{{\bar{p}}}
\def\brq{{\bar{q}}}

\def\brx{\bar{x}}
\def\bry{\bar{y}}

\def\brpartial{\bar{\partial}}

\def\brC{\bar{C}}

\def\brR{\bar{R}}

\def\brV{{\bar{V}}}

\def\brX{{\bar{X}}}
\def\brY{{\bar{Y}}}
\def\bX{{\bar{X}}}
\def\bY{{\bar{Y}}}
\def\brP{{\bar{P}}}

\newcommand{\na}{{\nabla}}

\newcommand{\bea}{\begin{eqnarray}}
\newcommand{\eea}{\end{eqnarray}}

\definecolor{rougef}{rgb}{0.56,0,0}
\definecolor{vertf}{rgb}{0,0.5,0}
\definecolor{bleuf}{rgb}{0,0,0.8}

\newcommand{\gammaws}{h}

\newcommand{\lA}{M}
\newcommand{\lB}{N}
\newcommand{\lC}{P}
\newcommand{\lD}{Q}

\newcommand{\Lie}{\mathcal{L}}

\begin{document}

%%%%%%%%%%%%%%%%%%%%%%%%%%%%%%%%%%%%%%%%%%%%%%%%%%%%%%%%%%%%%%%%%%%%%%%%%%%%%%%%
\begin{titlepage}

\title{
\bf Non-Riemannian isometries from double field theory\\
}

\author{ Chris\, D.  A.\,  Blair${}^{1}$\,,\quad Gerben Oling${}^{2}$\,,\quad and\,\quad Jeong-Hyuck Park${}^{3,4}$\footnote{On sabbatical leave from 3.}}
\date{}
\maketitle
\begin{center}
{\textit{
${}^{1}$ Theoretische Natuurkunde, Vrije Universiteit Brussel, and the International Solvay
Institutes, \\Pleinlaan 2, B-1050 Brussels, Belgium\\
\vspace{3pt}
${}^{2}$
The Niels Bohr Institute, University of Copenhagen,\\
Blegdamsvej 17, DK-2100 Copenhagen {\O}, Denmark \\
\vspace{3pt}
${}^{3}$ Department of Physics, Sogang University, 35 Baekbeom-ro, Mapo-gu,  Seoul  04107, Korea\\
${}^{4}$ Center for Gravitational Physics,
Yukawa Institute for Theoretical Physics, Kyoto University,\\
Kitashirakawa Oiwakecho, Sakyo-ku, Kyoto 606-8502, Japan\\
~\\}}
\texttt{christopher.blair@vub.be\qquad gerben.oling@nbi.ku.dk\qquad park@sogang.ac.kr }\\
~\\
~\\
\end{center}
\begin{abstract}
\noindent
We explore the notion of isometries in non-Riemannian geometries.
Such geometries include and generalise the backgrounds of non-relativistic string theory,
and they can be naturally described
using the formalism of double field theory.
Adopting this approach, we first solve the corresponding Killing equations for constant flat non-Riemannian backgrounds and show that they admit an infinite-dimensional algebra of isometries which includes a particular type of supertranslations.
These symmetries correspond to known worldsheet Noether symmetries of the Gomis--Ooguri non-relativistic string, which we now interpret as isometries of its  non-Riemannian doubled background.
We further consider the extension to supersymmetric double field theory and show that the corresponding Killing spinors can depend arbitrarily on the non-Riemannian directions, leading to ``supersupersymmetries'' that square to supertranslations.
\end{abstract}

~\\
~\\
\flushright{YITP-20-165}

\thispagestyle{empty}

\end{titlepage}

{\small{\tableofcontents}}

\section{Introduction}

The geometric interpretation of symmetries plays a key role in modern theoretical physics.
Perhaps the most prominent example of this is given by Lorentz symmetry, which was obscured in the original form of Maxwell's equations but is now at the heart of the unification of space and time in special and general relativity.
In special relativity, Lorentz symmetry corresponds to the coordinate transformations that are isometries of the (fixed) Minkowski metric.
In general relativity, the metric is dynamical and arbitrary coordinate transformations are allowed, but Lorentz transformations are distinguished as the (local) symmetries that are seen by a freely-falling observer.

However, the notion of general coordinate transformations and a dynamical metric with curvature is not tied to Lorentz symmetry.
Indeed, starting with the work of Cartan~\cite{Cartan1,Cartan2} it has been understood that the Galilean symmetries of non-relativistic physics can be realised as local symmetries of a covariant notion of geometry, known as Newton--Cartan geometry.
(See~\cite{Andringa:2010it,Christensen:2013lma,Hartong:2015zia} for the recent generalisation to torsional Newton--Cartan geometry.)
Likewise, the opposite ultra-relativistic limit (where the speed of light is taken to infinity) is associated to what is known as Carroll geometry~\cite{LevyLeblond,Duval:2014uoa,Ciambelli:2019lap}.
Such \emph{non-Riemannian} notions of geometry are of interest both as approximations of underlying relativistic theories and as interesting theories in their own right.

In the context of string theory, an important example of a  non-relativistic limit is given by the Gomis--Ooguri string~\cite{Gomis:2000bd,Danielsson:2000gi}, which can be obtained from a relativistic string in flat space using a limit that distinguishes two target space directions, with a compensating divergent $B$-field added to cancel the rest mass divergence of the string.
The resulting action is
\begin{equation}
  \label{eq:standard-gomis-ooguri-action}
  S_{\text{GO}}
  = \frac{T}{2} \int d^2\sigma\, \delta_{ab} \partial_- x^a \partial_+ x^b
  + \beta \partial_{-} y
  + \bar\beta \partial_{+} \bar{y}
  \,.
\end{equation}
This action is UV-finite and has a non-relativistic spectrum~\cite{Gomis:2000bd}.
It describes the motion of a string in $(D-2)$ flat target space directions $x^a$ with a Euclidean metric $\delta_{ab}$, together with two directions $y$ and $\bar{y}$ that (at first sight) do not obviously couple to a target space structure.
Instead, they couple to the two fields $\beta$ and $\bar\beta$, which can be viewed as Lagrange multipliers  constraining the embedding coordinates $y$ and $\bar y$ to be  chiral and anti-chiral respectively  on the worldsheet. This part of the action is exactly that of a $\beta \gamma$ system (or rather a chiral and anti-chiral pair of $\beta \gamma$ systems) \cite{Friedan:1985ey,Nekrasov:2005wg}.

While the conventional string action in  flat  background  is invariant under the finite-dimensional Poincar\'e isometries of the target spacetime, the non-relativistic action~\eqref{eq:standard-gomis-ooguri-action} is invariant not only under a form of target space Galilean symmetries, but in fact under an infinite-dimensional set of transformations~\cite{Batlle:2016iel,Bergshoeff:2019pij,Duval:1993pe}
\begin{equation}
  \label{eq:standard-gomis-ooguri-chiral-reparam}
  \delta x^a = \zeta^a(y),
  \quad
  \delta y = \zeta(y),
  \quad
  \delta \beta =
  - \frac{\partial \zeta(y)}{\partial y} \beta
  - 2 \frac{\partial \zeta^a(y)}{\partial y} \partial_+ x_a \,,
\end{equation}
and similarly for the $\bar y$ directions.
Here, $\zeta^a(y)$ and $\zeta(y)$ are arbitrary functions.
Hence the transformations of the coordinates are examples of the `supertranslational' symmetries familiar from the BMS asymptotic symmetry algebra of null infinity in asymptotically flat spacetimes~\cite{Bondi:1962px,Sachs:1962wk,Sachs:1962zza}.

In this paper, we will interpret the symmetries \eqref{eq:standard-gomis-ooguri-chiral-reparam} as target space \textit{isometries}, with the target space geometry provided by an enlarged geometry borrowed from the $\ODD$-invariant double field theory (DFT) \cite{Tseytlin:1990nb,Siegel:1993xq,Siegel:1993th,Gualtieri:2003dx,Hull:2004in,Hull:2009mi,Hohm:2010pp} description of string theory and supergravity.
This involves a `doubled' geometry usually motivated by a desire to geometrise the $\ODD$ T-duality symmetry of strings.\footnote{\label{SEP} The $\ODD$ symmetry also implies the doubling of the target space local Lorentz symmetry, $\mathbf{O}(1,D-1) \rightarrow \mathbf{O}(1,D-1)_L \times \mathbf{O}(1,D-1)_R$, with one factor seen by left-movers and the other by right-movers, see~\eqref{PbrPVbrV}.
  Note that we will often have in mind a spacetime dimension $D$ of 26 or 10.
}

In its usual Riemannian parametrisation, DFT combines the $D$-dimensional non-degenerate background metric $g_{\mu\nu}$ and the $B$-field $B_{\mu\nu}$ into a $2D$-dimensional generalised metric $\cH_{AB}$.
However, the latter can also be defined abstractly as follows.
We first introduce the $\ODD$-invariant metric~$\cJ_{AB}$, which with its inverse is used to lower and raise the $\ODD$ fundamental indices $A,B=1,\dots,2D$, and which takes an off-diagonal block form,
\be
\cJ_{AB} = \begin{pmatrix} 0 & 1 \\ 1 & 0 \end{pmatrix}\,.
\label{cJdef}
\ee
With this, the $\ODD$ tensor $\cH_{AB}$ is defined by the relations
\be
\ba{ll}
\cH_{AB}=\cH_{BA}\,,\qquad&\qquad
\cH_{A}{}^{C}\cH_{B}{}^{D}\cJ_{CD}=\cJ_{AB}\,.
\ea
\label{defcH}
\ee
Starting from the above, it is well known how a Riemannian metric $g_{\mu\nu}$ and the $B$-field $B_{\mu\nu}$ can be recovered \cite{Giveon:1988tt,Duff:1989tf}  ---see  \eqref{Hnormal}.
However, DFT works perfectly well given only the abstract definition of the generalised metric above, and thus is capable of going beyond the Riemannian paradigm.
This was initially noted empirically by acting with $\ODD$ rotations on known solutions as in \cite{Lee:2013hma,Ko:2015rha}.
By solving the conditions \eqref{defcH} in full generality, a unified description of non-Riemannian parametrisations was then classified in \cite{Morand:2017fnv}.
This was further studied in for instance \cite{Cho:2018alk,Cho:2019ofr,Park:2020ixf}.

As realised in \cite{Ko:2015rha}, where the Gomis--Ooguri action~\eqref{eq:standard-gomis-ooguri-action} was obtained from a non-Riemannian generalised metric via a doubled string worldsheet action, these parametrisations are relevant for the description of non-relativistic strings.
Entirely independently of this link, the latter field has seen much recent activity.
Building on earlier works~\cite{Gomis:2005pg,Andringa:2012uz} (see also~\cite{Gomis:2016zur}), the flat space theory~\eqref{eq:standard-gomis-ooguri-action} has been generalised to arbitrary stringy Newton--Cartan (SNC) backgrounds~\cite{Bergshoeff:2018yvt,Bergshoeff:2019pij}, which distinguish two spacetime directions corresponding to the $y$ and $\bar y$ above.
The action~\eqref{eq:standard-gomis-ooguri-action} can  be also obtained from a null reduction\footnote{An alternative higher-dimensional geometric interpretation of $\beta \gamma$ systems was recently studied in \cite{Lindstrom:2020oow}. Here null reductions play an important role, suggesting there should be a direct link between this work and the non-relativistic/non-Riemannian point of view.}
of a relativistic string, and for general backgrounds this results in a non-relativistic string coupling to torsional Newton--Cartan (TNC) backgrounds, which distinguishes a single time direction, plus an additional winding direction~\cite{Harmark:2017rpg,Harmark:2018cdl}.
The string beta functions have been computed for both the SNC~\cite{Gomis:2019zyu,Yan:2019xsf,Bergshoeff:2019pij} and TNC~\cite{Gallegos:2019icg} non-relativistic strings.
Furthermore, their actions have been identified on a classical level~\cite{Harmark:2019upf}, suggesting a duality between SNC and TNC non-relativistic string theories.

The DFT perspective on these models has been further developed in \cite{Park:2016sbw,Berman:2019izh, Blair:2019qwi,Cho:2019ofr} and has recently been comprehensively exploited in \cite{Gallegos:2020egk} to study the background field equations.
This makes the relationship between TNC and SNC more apparent, extending the identification of these non-relativistic string theories to the quantum level.%
\footnote{The necessity of and the relation between the various torsion constraints that were employed in the direct computations of the TNC~\cite{Gallegos:2019icg} and SNC~\cite{Gomis:2019zyu,Yan:2019xsf,Bergshoeff:2019pij} string beta functions still remain to be fully understood.
  We will come back to these questions and the relation to our work in Section~\ref{CONSection}.
}
\\

\noindent  In this work, we present a geometric derivation of the reparametrisation symmetry~\eqref{eq:standard-gomis-ooguri-chiral-reparam} and generalise it, together with the flat space   string action~\eqref{eq:standard-gomis-ooguri-action}, to arbitrary non-Riemannian backgrounds.
We first review the `doubled geometry' framework of $\ODD$ generalised geometry or double field theory with its non-Riemannian parametrisations in {\bf section~\ref{geoexamples}}.
The target spacetime local symmetries in DFT are $\ODD$ compatible generalised diffeomorphisms and lead to doubled Killing equations, which we present in detail  for non-Riemannian geometries in {\bf section~\ref{genKilling}}.

We will then show in {\bf section \ref{solvingKilling}} that the infinite-dimensional symmetry~\eqref{eq:standard-gomis-ooguri-chiral-reparam} is in fact the isometry of a  generalised metric.
For this, we shall  derive  all isometries of a generic constant  generalised metric by solving the relevant   Killing equations.
The solution is presented in equation~\eqref{KillingSOL}.
Besides, we discuss the separate Killing equation for the generalised dilaton in {\bf section \ref{furtherKilling}}.

We further consider the supersymmetric counterpart of these symmetries, relying on the  supersymmetric extensions of the $\ODD$ formalism~\cite{Siegel:1993th,Coimbra:2011nw,Jeon:2011vx,Jeon:2012hp}, in particular the maximally supersymmetric case of  \cite{Jeon:2012hp}, by solving the appropriate generalised Killing spinor equations in {\bf section \ref{KillingSpinor}}. As the most general solution  for the constant flat non-Riemannian background, we derive  supertranslational supersymmetries, or `supersupersymmetries' for short, see equation~\eqref{GSKS}. Further ---as can be expected from the usual results of supersymmetry, applied to the DFT case \cite{Jeon:2011sq,Hohm:2011nu}---  we verify that the commutator of supersupersymmetries generates precisely the supertranslational Killing vector  obtained  in  previous subsections.

We then  review the realisation of these non-Riemannian geometries as the target spacetime background  for a string, and  show how the infinite-dimensional isometry algebra leads to an infinite set of Noether symmetries on the worldsheet.
This is the focus of {\bf section \ref{StringSection}}. We conclude with various comments in {\bf section \ref{CONSection}}.

In {\bf Appendix \ref{derivKilling}} we describe our derivation of the most general Killing vector solution for the flat non-Riemannian background.
In {\bf Appendix \ref{AppendixCurved}} we carry out the same analysis for an example of a  curved non-Riemannian background, namely that considered already in \cite{Lee:2013hma}.

\section{Non-Riemannian geometries and their isometries
\label{isometries}}
In this section, after reviewing the appropriate notion of Lie derivatives in double field theory (DFT) as well as the Riemannian and non-Riemannian parametrisations of the DFT metric $\cH_{AB}$ and dilaton $d$, we solve the Killing equations for a generic  flat non-Riemannian background.
This gives rise to an infinite-dimensional set of isometries, which are algebraically similar to the supertranslations that arise in the BMS algebra of asymptotic symmetries of flat spacetime.
We continue to introduce and solve the corresponding Killing spinor equations on the same  flat non-Riemannian background, which leads to a supersymmetric analog of the supertranslations, or `supersupersymmetries'.

\subsection{Riemannian and non-Riemannian geometries
\label{geoexamples}}

The local symmetries compatible with the existence of the $\ODD$-invariant metric~\eqref{cJdef} are \emph{generalised diffeomorphisms}.
They are generated by generalised vectors $\Lambda^A$ via the generalised Lie derivative $\hcL$, which acts on a generalised tensor density $T_{A_1\cdots A_n}$, with weight $\omega$, as
\begin{equation}
  \hcL_\Lambda T_{A_1 \cdots A_n}
  = \Lambda^B \partial_B T_{A_1\cdots A_n}+ \omega \partial_{B}\Lambda^{B}T_{A_{1}\cdots A_{n}}
  + \sum_{j=1}^n~ 2 \partial_{[A_j} \Lambda_{B]} T_{A_1 \cdots A_{j-1}}{}^B{}_{A_{j+1} \cdots A_n}
  \,.
  \label{GLD}
\end{equation}
This definition ensures that $\hcL_{\Lambda} \cJ_{AB} = 0$, and thus the fundamental  $\ODD$ structure is preserved.
We mention again that indices are raised and lowered using the
metric $\cJ_{AB}$ and its inverse.
The principal $\ODD$ tensors that we will encounter are the generalised metric, $\cH_{AB}$, which is weightless,  and the generalised dilaton, $d$, for which the exponential $e^{-2d}$ has  weight  one and provides  the integral measure in DFT.

In \eqref{GLD}, we have partial derivatives $\partial_A$ with respect to a set of $2D$-dimensional coordinates $x^A$.
However, the actual coordinate dependence is constrained by the \emph{section condition}:
\be
\cJ^{AB} \partial_A \partial_B \mathcal{O} = 0\,,\qquad
\cJ^{AB} \partial_A \mathcal{O} \partial_B \mathcal{O}^\prime = 0\,,
\label{sectioncondition}
\ee
where $\mathcal{O},\mathcal{O}^\prime$ stands for any field or gauge parameter. This ensures that at most half the coordinates are ``physical'', and
 guarantees that the generalised Lie derivatives form a closed algebra,
\be
\left[\hcL_{\Lambda_{1}},\hcL_{\Lambda_{2}}\right]=\hcL_{\left[\Lambda_{1},\Lambda_{2}\right]_{\rm{C}}}\,,
\ee
with the following anti-symmetric bracket
\be
\left[\Lambda_{1},\Lambda_{2}\right]^{A}_{\rm{C}}
\equiv\half\!\left(\hcL_{\Lambda_{1}}\Lambda_{2}^{A}-\hcL_{\Lambda_{2}}\Lambda_{1}^{A}\right)
=\Lambda_{1}^{B}\partial_{B}\Lambda_{2}^{A}-\Lambda_{2}^{B}\partial_{B}\Lambda_{1}^{A}+\half \Lambda_{2}^{B}\partial^{A}\Lambda_{1B}-\half \Lambda_{1}^{B}\partial^{A}\Lambda_{2B}
\,.
\label{Cbracket}
\ee
Note that the generalised Lie derivative on vector fields itself is \emph{not} anti-symmetric,
\be
\hcL_{\Lambda_1} \Lambda_2^A = \Lambda_1^B \partial_B \Lambda_2^A - \Lambda_2^B \partial_B \Lambda_1^A + \Lambda_2^B \partial^A \Lambda_{1B}\,,
\ee
but the bracket~\eqref{Cbracket} that arises from its action on other tensor fields \emph{is}  so.

The connection between this doubled formalism and the usual spacetime picture based on $\mathbf{GL}(D)$ diffeomorphisms comes about as follows. In view of the form of the  $\ODD$-invariant metric \eqref{cJdef}, we  can decompose the doubled coordinates as $x^A = (\tilde x_\mu, x^\nu)$, where $\mu,\nu,\dots$ denote $D$-dimensional coordinate indices.
Writing out the section condition in this decomposition, $\partial_{A}\partial^{A}=\partial_{\mu}\tpartial^{\mu}+\tpartial^{\mu}\partial_{\mu}=0$, the conventional solution is to take the field to only depend on the $x^\mu$, i.e. ${\tilde \partial^\mu \equiv 0}$ acting on anything.
Writing the generalised vector $\Lambda^A = (\lambda_\mu, \xi^\nu)$, the generalised Lie derivative then describes the usual Lie derivative, with respect to the vectorial part, $\Lambda^\mu \equiv \xi^\mu$,  plus gauge transformations of a two-form potential $B$-field, corresponding to the one-form part, $\Lambda_\nu \equiv \lambda_\nu$.

Alongside this split of coordinates, we have to choose how we parametrise the generalised metric~$\cH_{AB}$ in terms of fields carrying $D$-dimensional indices, such that the defining conditions~\eqref{defcH} are satisfied.
Usually, one is led to a $D$-dimensional (pseudo-)Riemannian parametrisation which we briefly review below.
However, this parametrisation is not the most general one: the geometric data of DFT, i.e.~$\cH_{AB}$ and $d$,  can in fact also be described through  non-Riemannian variables.
We shall illustrate this for the parametrisation corresponding to `stringy' Newton--Cartan geometry and then discuss the most general cases.

\subsubsection*{Riemannian geometry}

The following parametrisation of the generalised metric corresponds to a standard string theory background, in terms of a Riemannian (or Lorentzian) metric $g_{\mu\nu}$ and a two-form $B_{\mu\nu}$~\cite{Giveon:1988tt,Duff:1989tf}:
\be
\cH_{AB} = \begin{pmatrix}
g^{\mu\nu} & - g^{\mu \rho }B_{\rho \nu} \\
B_{\mu\rho} g^{\rho \nu} & g_{\mu\nu} - B_{\mu\rho} g^{\rho\sigma} B_{\sigma \nu}
\end{pmatrix}\,.
\label{Hnormal}
\ee
In addition to $\cH_{AB}$, the NS-NS sector is described by the generalised dilaton $d$, which is related to the usual dilaton $\phi$ by
\be
e^{-2d} = e^{-2\phi} \sqrt{|g|}\,.
\label{gendil}
\ee
As mentioned previously, and as is clear here,
$e^{-2d}$ has unit weight, implying its transformation under generalised diffeomorphisms to read $\hcL_{\Lambda}d\equiv-\half e^{2d}\hcL_{\Lambda}\big(e^{-2d}\big)=\Lambda^{A}\partial_{A}d-\half\partial_{A}\Lambda^{A}$.
These two fields provide the starting point for discussing $\ODD$ generalised geometry or doubled formulations of string theory and supergravity~\cite{Tseytlin:1990nb,Siegel:1993xq,Siegel:1993th,Gualtieri:2003dx,Hull:2004in,Hull:2009mi,Hohm:2010pp}.

\subsubsection*{Non-relativistic geometry (stringy Newton--Cartan)}

An illustrative example of a non-Riemannian geometry comes from the `stringy' non-relativistic limit of a Riemannian geometry leading to stringy Newton--Cartan (SNC) geometry~\cite{Andringa:2012uz,Bergshoeff:2018yvt,Bergshoeff:2019pij,Harmark:2019upf}.
This geometry can be constructed using an expansion of the Riemannian variables in the parametrisation~\eqref{Hnormal} in powers of $1/c^2$.
In this expansion, a time-like and a space-like direction aligned with the string worldsheet are distinguished, and a divergent term in the $B$-field flux is used to compensate the divergence in the Riemannian metric as $c\to\infty$.
As a result, in the limit, one ends up with a finite string action and a well-defined notion of SNC geometry.

In double field theory, the SNC limit can be carried out directly in terms of the Riemannian parametrisation~\eqref{Hnormal} of the generalised metric.
Exactly this limit was analysed in Section 3.3 of~\cite{Morand:2017fnv}; here we make an explicit connection to the usual SNC variables.
We use $\lA, \lB$ to label the ``longitudinal'' directions along the worldsheet, so that $\eta_{MN}$ denotes the two-dimensional flat metric (with which we will raise and lower longitudinal indices below) and $\epsilon_{MN}$ is the two-dimensional alternating symbol.
With that, we perform the following expansion of the metric, inverse metric and $B$-field:
\begin{align}
\label{SNCexp}
g_{\mu \nu}& = c^2 \eta_{\lA \lB} \tau_{\mu}\,^{\lA} \tau_{\nu}\,^{\lB} + H^{\perp}_{\mu \nu}+\eta_{\lA \lB} \tau_{\mu}\,^{\lA} m_{\nu}\,^{\lB}+\eta_{\lA \lB} \tau_{\nu}\,^{\lA} m_{\mu}\,^{\lB}
+ O(c^{-2})
\,,\\\nonumber
g^{\mu \nu} &= H^{\perp}{}^{\mu \nu}+{c}^{-2} \left(v^{\mu}\,_{\lA}-H^{\perp \mu \rho} m_{\rho}\,^{\lC} \eta_{\lA \lC}\right) \left(v^{\nu}\,_{\lB}-H^{\perp \nu \sigma} m_{\sigma}\,^{\lD} \eta_{\lB \lD}\right) \eta^{\lA \lB}
+ c^{-4} Y^{\mu \nu} +O(c^{-6})\,,\\\nonumber
B_{\mu \nu} & = -c^2 \epsilon_{\lA \lB} \left(\tau_{\mu}\,^{\lA}+{c}^{-2} m_{\mu}\,^{\lA}\right) \left(\tau_{\nu}\,^{\lB}+{c}^{-2} m_{\nu}\,^{\lB}\right)+\bar B_{\mu \nu}
+O(c^{-2})\,.
\end{align}
Here, we have parametrised the expansion using the longitudinal vielbein $\tau_\mu{}^\lA$ and the degenerate ``transverse'' metric $H_{\mu\nu}^\perp$, along with their projective inverses $v^\mu{}_\lA$ and $H^{\perp \mu \nu}$, which satisfy
\be
\tau_\mu{}^\lA v^\mu{}_\lB = \delta^\lA{}_\lB \,,\quad
\tau_\mu{}^\lA H^\perp{}^{\mu\nu} = 0\,,\quad
v^\mu{}_\lA H^\perp{}_{\mu\nu} = 0\,,\quad
H^{\perp}{}^{\mu \rho} H^\perp_{\rho \nu} + v^\mu{}_\lA \tau_{\nu}{}^\lA = \delta^\mu{}_\nu\,.
\label{SNCcomplete}
\ee
In addition, $m_\mu{}^\lA$ is the stringy Newton--Cartan one-form, and $\bar B_{\mu\nu}$ are the subleading components of the $B$-field.
Note that we need to expand the Riemannian metric as well as its inverse, since they both appear in the generalised metric~\eqref{Hnormal}.
However, the subleading components in the expansion of $g^{\mu\nu}$ can be fixed using
the expansion of the identity $g_{\mu\rho} g^{\rho \nu} = \delta_\mu{}^\nu$.
In particular, from this expansion we can extract the longitudinal part of the quantity~$Y^{\mu\nu}$,
\be
Y^{\mu\nu} \tau_{\mu}{}^{\lA} \tau_{\nu}{}^{\lB} = - 2 m_{\mu}{}^{ (\lA} v^{|\mu|}{}^{\lB)} + m_{\rho}{}^{\lA} H^{\rho \sigma} m_{\sigma}{}^{ \lB} \,.
\label{Ylong}
\ee
We now consider the non-relativistic $c\to\infty$ limit of the Riemannian parametrisation~\eqref{Hnormal} of the generalised metric using the expansion~\eqref{SNCexp}.
All components are finite in this limit, and after using \eqref{Ylong} we can write them as
\be
\mathcal{H}_{AB} =
\begin{pmatrix}
H^{\perp}{}^{\mu \nu} & - H^\perp{}^{\mu \rho} \bar B_{\rho \nu} + \epsilon_{\lA \lB} v^{\mu \lA} \tau_{\nu}{}^{\lB} \\
\epsilon_{\lA\lB} v^{\nu \lA}\tau_{\mu}{}^{\lB}+\bar B_{\mu \rho} H^\perp{}^{\rho \nu} & H^\perp_{\mu \nu}- \bar B_{\mu \rho}  H^{\perp \rho \sigma} \bar B_{\sigma \nu }
+ 2 \epsilon_{\lA \lB} v^{\rho \lA}\tau_{(\mu}\,^{\lB} \bar B_{\nu) \rho}
\end{pmatrix} \,.
\label{HSNC}
\ee
Using the relations~\eqref{SNCcomplete}, we now see  that the upper left block $\cH^{\mu\nu}=H^{\perp}{}^{\mu \nu}$ is non-invertible.
As such, we can no longer identify this block with the inverse Riemannian spacetime metric, as in the Riemannian parametrisation~\eqref{Hnormal}.
Instead, equation~\eqref{HSNC} parametrises the generalised metric $\cH_{AB}$ in terms of a stringy Newton--Cartan geometry, which is defined by the inverse transverse metric $H^\perp{}^{\mu\nu}$ and its longitudinal zero vectors $\tau_\mu{}^\lA$.
The latter can be used to define an additional degenerate metric $\tau_{\mu\nu}\equiv \tau_{\mu}{}^{\lA} \tau_{\nu}{}^{\lB} \eta_{\lA \lB}$ on the longitudinal directions.
Note that the one-form $m_\mu{}^\lA$ drops out of the generalised metric completely, in agreement with the analysis of \cite{Morand:2017fnv,Harmark:2019upf}.
The SNC parametrisation can be gauge fixed to recover the TNC parametrisation~\cite{Harmark:2019upf} (the direct embedding of TNC into DFT was formulated in \cite{Berman:2019izh, Blair:2019qwi}).

We can then also consider the generalised dilaton, $e^{-2d}$, which starting from \eqref{gendil} in the Riemannian case is finite in the limit assuming an appropriate expansion of the usual dilaton $\phi= \bar\phi+\ln c$ \cite{Bergshoeff:2019pij}, leading to
\be
e^{-2d} = e^{-2\bar\phi} \sqrt{\det{}^\prime H^{\perp}} \,,\quad
\det{}^\prime H^{\perp} \equiv \frac{\epsilon^{\mu_1 \dots \mu_D} \epsilon^{\nu_1\dots \nu_D}}{(D-2)!}  \tau_{\mu_1}^M \tau_{\mu_2}^N \tau_{\nu_1}^P \tau_{\nu_2}^Q \epsilon_{MN} \epsilon_{PQ}
H^\perp_{\mu_3\nu_3} \dots H^\perp_{\mu_D\nu_D}\,.
\label{SNCgendil}
\ee
The generalised metric \eqref{HSNC} and dilaton \eqref{SNCgendil} can then be inserted into the action and equations of motion of DFT\footnote{In the absence of other fields, the equations of motion of  $\cH_{AB}$ and $d$ are the vanishing of the generalised Ricci tensor and generalised Ricci scalar \cite{Siegel:1993th}. Unlike in general relativity, these are two independent equations, however they can be combined by defining a generalised Einstein tensor which obeys a generalised Bianchi identity and unifies the field equations into a single familiar  form, $G_{AB} = 8 \pi G T_{AB}$~\cite{Angus:2018mep}.} as well as the doubled string actions.
The  equations of motion of DFT can also be obtained \cite{Berman:2007xn,Copland:2011yh} as the beta functionals for the doubled sigma models we consider below (which reproduce those of non-relativistic strings).
In particular, this means that they should be independent of the parametrisation of the generalised metric.
The subtlety that arises, as analysed for general non-Riemannian backgrounds in \cite{Cho:2019ofr}, is that if one wishes to maintain the non-relativistic parametrisation of the generalised metric~\eqref{HSNC},   one must forbid certain variations (those which would make the block $\cH^{\mu\nu}$ again invertible).
For SNC, this restriction would lead to one fewer equation of motion than expected from DFT.
A detailed analysis and the comparison to the non-relativistic beta-functional equations \cite{Gomis:2019zyu,Gallegos:2019icg,Bergshoeff:2019pij,Yan:2019xsf} has now been carried out in detail by Gallegos, Gürsoy, Verma and Zinnato~\cite{Gallegos:2020egk}.

\subsubsection*{General non-Riemannian geometries}

The general classification of non-Riemannian geometries in DFT, carried out in \cite{Morand:2017fnv}, allows $\cH^{\mu\nu}$ to have arbitrary rank.
Let $\cH^{\mu\nu} = H^{\mu\nu}$, where $H^{\mu\nu}$ is a symmetric (possibly) degenerate matrix, further introduce $K_{\mu\nu}$, likewise symmetric and (possibly) degenerate, and let a basis for the kernels of $H$ and $K$ be given by
$\big\{X^{i}_{\mu},\brX^{\bri}_{\nu}\big\}$  and $\big\{Y_{j}^{\mu},\brY^{\nu}_{\brj}\big\}$ respectively, such that with $i,j=1,2,\ldots, n\,$ and  $\,\bri,\brj=1,2,\ldots,  \brn$ we have
\be
H^{\mu\nu}X^{i}_{\nu}=0\,,\qquad
H^{\mu\nu}\brX^{\bri}_{\nu}=0\,,\qquad
K_{\mu\nu}Y_{j}^{\nu}=0\,,\qquad
K_{\mu\nu}\brY_{\brj}^{\nu}=0\,.
\label{HXX}
\ee
These obey the following completeness relation,
\be
H^{\mu\rho}K_{\rho\nu}
+Y_{i}^{\mu}X^{i}_{\nu}+\brY_{\bri}^{\mu}\brX^{\bri}_{\nu}
=\delta^{\mu}{}_{\nu}\,,
\label{COMP}
\ee
implying
\be
\begin{aligned}
& Y^{\mu}_{i}X_{\mu}^{j}=\delta_{i}{}^{j}\,,\quad
& \brY^{\mu}_{\bri}\brX_{\mu}^{\brj}=\delta_{\bri}{}^{\brj}\,,\qquad
& H^{\rho\mu}K_{\mu\nu}H^{\nu\sigma} =H^{\rho\sigma}\,,
\\
& Y^{\mu}_{i}\brX_{\mu}^{\brj}= 0\,,\quad
& \brY^{\mu}_{\bri}X_{\mu}^{j}=0\,,\qquad
& K_{\rho\mu}H^{\mu\nu}K_{\nu\sigma}=K_{\rho\sigma}\,,
\label{complete}
\end{aligned}
\ee
and allow us to write the generalised metric as
\be
\cH_{AB}=
\begin{pmatrix}
H^{\mu\nu}&
-H^{\mu\rho}B_{\rho\nu}+Y_{i}^{\mu}X^{i}_{\nu}-
\brY_{\bri}^{\mu}\brX^{\bri}_{\nu}\\
B_{\mu\rho}H^{\rho\nu}+X^{i}_{\mu}Y_{i}^{\nu}
-\brX^{\bri}_{\mu}\brY_{\bri}^{\nu}
\,\,
&K_{\mu\nu}-B_{\mu\rho}H^{\rho\sigma}B_{\sigma\nu}
+2X^{i}_{(\mu}B_{\nu)\rho}Y_{i}^{\rho}
-2\brX^{\bri}_{(\mu}B_{\nu)\rho}\brY_{\bri}^{\rho}
\end{pmatrix}
\,,
\label{cHFINAL}
\ee
such that the $\ODD$ compatibility condition \eqref{defcH} is indeed satisfied.
Here, as usual, the $B$-field is skew-symmetric.
The $\ODD$ invariant trace is $\cJ^{AB} \cH_{AB} = 2(n-\brn)$.
The generalised metric \eqref{cHFINAL} parametrises an underlying coset~\cite{Berman:2019izh} given by $\frac{\mathbf{O}(D,D)}{\mathbf{O}(t+n,s+n)\times\mathbf{O}(s+\brn,t+\brn)}$, where the signature $(-,+,0)$ of  both $H^{\mu\nu}$ and $K_{\mu\nu}$ is commonly  $(t,s,n+\brn)$.\footnote{This sets the signature of the twofold spin groups in DFT to be quite general, not necessarily Minkowskian, c.f.~\eqref{etas}.}
This coset has dimension $D^{2}-(n-\brn)^{2}$, which matches  the number of degrees of freedom in the infinitesimal  fluctuations (i.e.~moduli) around the above  $(n,\brn)$ background~\cite{Cho:2019ofr}.

The usual Riemannian case is included as $n=\brn=0$.
Stringy non-relativistic geometry has $n=\brn=1$.
Comparing \eqref{cHFINAL} with \eqref{HSNC}, one can immediately identify $K_{\mu\nu} = H^\perp_{\mu\nu}$, $H^{\mu\nu} = H^{\perp \mu\nu}$, $B_{\mu\nu} = \bar B_{\mu\nu}$ and $\tau_\mu{}^\lA$, $v^\mu{}_\lA$ with the zero vectors. For instance, using light-cone coordinates $\lA = (+,-)$ we can write $\tau_\mu{}^{\lA} = ( X_\mu, \bX_\mu )$, $v^\mu{}_{\lA} = ( Y^\mu, \bY^\mu)$ with $\epsilon_{MN}$ and $\eta_{MN}$ defined by $\epsilon_{+-} = -1$, $\eta_{+-} = 1$, $\eta_{++}=\eta_{--}=0$.
Other examples studied in \cite{Morand:2017fnv} include a version of Carroll geometry with $n=D-1$, $\brn=0$.
However, note that the (BRST) quantum consistency of the doubled string appears to impose $n=\brn$~\cite{Park:2020ixf}.

The generalised dilaton, generalising \eqref{SNCgendil}, can be written as
\be
e^{-2d} = e^{-2 \phi} \sqrt{|\det{}^\prime K|}\,,
\label{nonriegendil}
\ee
where
\be
\begin{split}
\det{}^\prime K & \equiv
\frac{\epsilon_{i_1 \dots i_n} \epsilon_{\bri_1 \dots \bri_n} \epsilon^{\mu_1 \dots \mu_D}
\epsilon_{j_1 \dots j_n} \epsilon_{\brj_1 \dots \brj_{\brn}} \epsilon^{\nu_1 \dots \nu_D}
}{(n! \brn! )^2 (D-n-\brn)!}
X_{\mu_1}^{i_1} \dots X_{\mu_n}^{i_n}
\bar X_{\mu_{n+1}}^{\bri_1} \dots \bar X_{\mu_{n+\brn}}^{\bri_{\brn}}
\\ & \qquad \times
X_{\nu_1}^{j_1} \dots X_{\nu_n}^{j_n}
\bar X_{\nu_{n+1}}^{\brj_1} \dots \bar X_{\nu_{n+\brn}}^{\brj_{\brn}}
K_{\mu_{n+\brn+1} \nu_{n+\brn+1}} \dots K_{\mu_D \nu_D} \,.
\end{split}
\label{uglydeterminant}
\ee
At this point, we would like to emphasise that, from the point of view of the Riemannian parametrisation, such non-Riemannian geometries are singular.
However, in the DFT formulation, we can describe them without any problems using the appropriate non-Riemannian parametrisation of the generalised metric.

\subsubsection*{Local Lorentz symmetries: $\GL(n)\times\GL(\brn)$  and Milne-shift}

The choice of  non-Riemannian parametrisation of the generalised metric \eqref{cHFINAL} is not completely rigid.
Two parts of the underlying local Lorentz symmetries,
$\mathbf{O}(t+n,s+n)\times\mathbf{O}(s+\brn,t+\brn)$, can be seen directly as transformations of the $D$-dimensional variables  appearing in the  parametrisation of the generalised metric~\eqref{cHFINAL}.
These are  $\GL(n)\times\GL(\brn)$ rotations and Milne-shift transformations.
Specifically,   the  $\GL(n)\times\GL(\brn)$ symmetry acts on    the unbarred $i,j,\cdots$ and barred $\bri,\brj,\cdots$ indices. On the other hand,  the  Milne-shift symmetry generalises  the   `Galilean  boost' in the Newtonian  gravity literature~\cite{Milne:1934,Duval:1993pe} and acts with local parameters, $V_{\mu i}$ and  $\brV_{\mu\bri}$,  as~\cite{Morand:2017fnv}
\begin{align}
\label{MS}
Y^{\mu}_{i}&\rightarrow Y^{\mu}_{i} + H^{\mu\nu}V_{\nu i}\,,\quad\qquad
\brY_{\bri}^{\mu}\rightarrow \brY^{\mu}_{\bri} + H^{\mu\nu}\brV_{\nu\bri}\,,\\
\nonumber
 K_{\mu\nu}&\rightarrow K_{\mu\nu} -2X^{i}_{(\mu}K_{\nu)\rho}H^{\rho\sigma}V_{\sigma i}-2\brX^{\bri}_{(\mu}K_{\nu)\rho}H^{\rho\sigma}\brV_{\sigma\bri}
+(X_\mu^iV_{\rho i} + \brX^{\bri}_\mu \brV_{\rho \bri} ) H^{\rho\sigma} (X_\nu^iV_{\sigma i} + \brX^{\bri}_\nu \brV_{\sigma \bri} )
\,,\\
\nonumber
B_{\mu\nu}&\rightarrow B_{\mu\nu}
-2X^{i}_{[\mu}V_{\nu]i}+2\brX^{\bri}_{[\mu}\brV_{\nu]\bri}
+2X^{i}_{[\mu}\brX^{\bri}_{\nu]}\left(Y_{i}^{\rho}\brV_{\rho\bri}
+\brY_{\bri}^{\rho}V_{\rho i}
+ V_{\rho i} H^{\rho \sigma} \brV_{\sigma \bri}
\right),
\end{align}
while leaving $H^{\mu\nu}$, $X_\mu^i$ and $\bar X_\mu^{\bri}$ invariant.
The generalised dilaton is also invariant: note then that the determinant, $\det{}^\prime K$, appearing in \eqref{nonriegendil} is invariant under Milne-shift transformations, but not under $\GL(n) \times \GL(\brn)$ transformations, hence (as the generalised dilaton $d$ does not transform) the scalar $\phi$ transforms under the latter.
From the perspective of double field theory, it is better to think of $d$ as more fundamental than $\phi$.
Note that \eqref{MS} is a finite transformation, and the infinitesimal variation, which we  henceforth denote by $\deltaM$, amounts to the terms linear in the local parameters.  The exponentiation, $e^{\deltaM}$, truncates at most  at the quadratic order as above.
These shift symmetries will play a minor role below.

\subsubsection*{Constant flat non-Riemannian geometry}

The simplest $(1,1)$ non-relativistic geometry~\cite{Ko:2015rha} we can consider is that obtained as in \cite{Gomis:2000bd, Danielsson:2000gi} by taking the SNC $c^2 \rightarrow \infty$ limit in flat $D$-dimensional spacetime.
An $(n,\brn)$ extended version of this geometry is given by the following generalised metric:
\be
\cH_{AB} =
\begin{pmatrix}
\eta^{ab}&0&0&0&0&0\\
0&0&0&0&\delta^{i}{}_{j}&0\\
0&0&0&0&0&-\delta^{\bri}{}_{\brj}\\
0&0&0&\eta_{cd}&0&0\\
0&\delta_{k}{}^{l}&0&0&0&0\\
0&0&-\delta_{\brk}{}^{\brl}&0&0&0
\end{pmatrix} \,,
\label{mcH0}
\ee
where $\eta_{ab}$ is the flat (Minkowski) metric of signature $(t,s)$ and we have chosen our coordinates to align with the zero vector directions, thus
\be
x^\mu = ( x^a, x^i, \bar x^{\bri} ) \,,\quad
a = 1,\dots,D-n-\brn \,,\quad
i = 1,\dots ,n\,,\quad
\bri = 1,\dots, \brn\,.
\label{flatcoords}
\ee
The `natural' choice for the parameterising  fields reads, up to $\GL(n)\times\GL(\brn)$ and Milne-shifts,
\be
H^{\mu\nu} = \begin{pmatrix} \eta^{ab} & 0 & 0	\\ 0 & 0 & 0 \\ 0 & 0 & 0 \end{pmatrix} \,,\qquad\quad
K_{\mu\nu} = \begin{pmatrix} \eta_{ab} & 0 & 0	\\ 0 & 0 & 0 \\ 0 & 0 & 0 \end{pmatrix} \,,\qquad\quad
B_{\mu\nu} = 0 \,,
\label{flatH_HKB}
\ee
\be\!\!
X_\mu^j =\delta_{\mu}^{j}
 \,,\qquad
\brX_\mu^{\brj} =\delta_{\mu}^{\brj}
\,,\qquad
Y^\mu_j=\delta^{\mu}_{j}
\,,\qquad
\brY^\mu_{\brj} =\delta^{\mu}_{\brj}
\,.
\label{flatH_XY}
\ee
The above  generalised metric~\eqref{mcH0}
is the natural candidate for ``flat''\footnote{The doubled formulation does not admit a well-defined notion of a Riemann tensor~\cite{Jeon:2010rw,Jeon:2011cn,Hohm:2011si}: the only unambiguous curvature tensors are those appearing in the equations of motion, which vanish if we restrict to the simplest matter-free bosonic DFT.
We call the geometry~\eqref{mcH0} flat since it is constant, hence all curvatures vanish, and it reduces to the (formally doubled description of) Minkowski spacetime for $n=\brn=0$.}
  non-Riemannian space.
We will next show how this geometry admits an infinite-dimensional family of isometries.

\subsection{Generalised metric Killing equation}
\label{genKilling}

The Killing equation for the generalised metric follows from requiring that its generalised Lie derivative with respect to some particular generalised vector $\Lambda$ vanishes:
\be
\hcL_{\Lambda}\cH_{AB}=
\Lambda^{C}\partial_{C}\cH_{AB}+2\partial_{[A}\Lambda_{C]}\cH^{C}{}_{B}
+2\partial_{[B}\Lambda_{C]}\cH_{A}{}^{C}=0\,.
\label{Killing}
\ee
For $\Lambda^{M}=(\lambda_{\mu},\xi^{\nu})$ and the section condition solution ${\tpartial^{\mu}\equiv0}$, the equation \eqref{Killing} consists of three parts:
\begin{equation}
  \label{genKillingGeneral}
  \begin{gathered}
    \hcL_\Lambda \cH^{\mu\nu} = \Lie_{\xi}\cH^{\mu\nu}\,,
    \qquad
    \hcL_\Lambda \cH_\mu{}^\nu =\Lie_{\xi}\cH_{\mu}{}^{\nu}+2\partial_{[\mu} \lambda_{\rho]}\cH^{\rho\nu} \,,
    \\
    \hcL_\Lambda \cH_{\mu\nu} =\Lie_{\xi}\cH_{\mu\nu}+2\partial_{[\mu} \lambda_{\rho]}\cH^{\rho}{}_{\nu}+ 2\partial_{[\nu} \lambda_{\rho]} \cH_\mu{}^\rho \,,
  \end{gathered}
\end{equation}
where $\Lie_\xi$ denotes the usual $D$-dimensional Lie derivative. Setting these  all equal to zero for the generic  $(n,\brn)$ parametrisation~\eqref{cHFINAL}, one finds the  Killing equations can be written as
\be
\begin{split}
0 & = \Lie_\xi H^{\mu\nu} \,,\\
0 & =(\Lie_\xi B_{\mu \rho} + 2 \partial_{[\mu} \lambda_{\rho]} ) H^{\rho \nu} + \Lie_\xi  (X_\mu^i Y_i^\nu -\bX_\mu^{\bri} \bY_{\bri}^\nu)\,,\\
0 & = \Lie_\xi K_{\mu\nu} + ( \Lie_\xi B_{\mu \rho} + 2 \partial_{[\mu} \lambda_{\rho]} )  (X_\nu^i Y_i^\rho -\bX_\nu^{\bri} \bY_{\bri}^\rho)
+ ( \Lie_\xi B_{\nu \rho} + 2 \partial_{[\nu} \lambda_{\rho]} )  (X_\mu^i Y_i^\rho -\bX_\mu^{\bri} \bY_{\bri}^\rho)\,.
\end{split}
\label{genKillingGeneral1}
\ee
As the generalised metric is constrained by the relations \eqref{HXX}, \eqref{COMP} and  \eqref{complete} obeyed by the fields appearing in the parametrisation, not all of the equations in~\eqref{genKillingGeneral1} are independent.
By taking projections with $\big\{K_{\mu\rho} H^{\rho \nu}, X_\mu^i Y_i^\nu,\bX_\mu^{\bri} \bY_{\bri}^\nu\big\}$,
we can obtain the following minimal set of Killing equations which are equivalent to \eqref{genKillingGeneral1} (and collectively $\GL(n)\times\GL(\brn)$, Milne-shift invariant):
\be
H^{\mu\nu} \Lie_\xi X_\nu^i = 0=H^{\mu\nu} \Lie_\xi \bX_\nu^{\bri} \,,\qquad
 \bY^\mu_{\bri} \Lie_\xi X_\mu^i = 0= Y^\mu_{i} \Lie_\xi \bX_\mu^{\bri} \,,\quad
\label{HLX}
\ee
\be
H^{\mu\rho} H^{\nu\sigma} \Lie_\xi K_{\rho\sigma} = 0 \,,\quad
H^{\mu\rho} H^{\nu\sigma} (\Lie_\xi B_{\rho \sigma} + 2 \partial_{[\rho} \lambda_{\sigma]} )=0\,,\quad
Y^\mu_i \bY^\nu_{\bri} ( \Lie_\xi B_{\mu\nu} + 2 \partial_{[\mu} \lambda_{\nu]} ) = 0\,,
\label{YLX}
\ee
\vspace{-1.5\baselineskip}
\begin{align}
Y^\rho_i H^{\mu \sigma}(\Lie_\xi B_{\rho\sigma} + 2 \partial_{[\rho} \lambda_{\sigma]} )+ H^{\mu\rho} K_{\rho\sigma} \Lie_\xi Y^\sigma_i
&= 0\,,\\
\bY^\rho_{\bri} H^{\mu \sigma}(\Lie_\xi B_{\rho\sigma} + 2 \partial_{[\rho} \lambda_{\sigma]} )- H^{\mu\rho} K_{\rho\sigma} \Lie_\xi \bY^\sigma_{\bri}
&=0 \,.
\end{align}
In the Riemannian case $(n,\brn)=(0,0)$, where $K_{\mu\nu}=g_{\mu\nu}$ is an invertible metric, these imply the usual conditions that $\Lie_\xi g_{\mu\nu}=0$ and $\Lie_\xi B_{\mu\nu}+2\partial_{[\mu} \lambda_{\nu]}=0$.
In the non-Riemannian case, such expressions would not be Milne-shift invariant and thus should not be satisfied identically.
Instead, certain of their contractions with $H^{\mu\nu}$, $Y^\mu_i$, $\bY^\mu_{\bri}$ are constrained as specified above.
(This can be thought of as projecting into different combinations of Riemannian and non-Riemannian chiral or anti-chiral directions.
Note the projections in the $Y^\mu_i Y^\nu_{j}$ and $\bY^\mu_{\bri} \bY^\nu_{\brj}$ directions of the $B$-field equation are completely unconstrained.)
Similarly, the projection of the variation $\Lie_\xi X_\mu^i$ must vanish except in the $Y^\mu_j$ directions, and so on.

If we define $\xi^i \equiv \xi^\mu X_\mu^i$ and $\bar \xi^{\bri} \equiv \xi^\mu \bX_\mu^{\bri}$ we can write \eqref{HLX} as
\be
\ba{ll}
H^{\mu\nu} \partial_\nu \xi^i = H^{\mu\nu} \xi^\rho ( \partial_\nu X_\rho^i - \partial_\rho X_\nu^i) \,,\quad&\quad
H^{\mu\nu} \partial_\nu \brxi^{\bri} = H^{\mu\nu} \xi^\rho ( \partial_\nu \bX_\rho^{\bri} - \partial_\rho \bX_\nu^{\bri}) \,,\\
\bY^{\mu}_{\bri} \partial_\mu \xi^i = \bY^{\mu}_{\bri} \xi^\nu ( \partial_\mu X_\nu^i - \partial_\nu X_\mu^i) \,,\quad&\quad
Y^\mu_i \partial_\mu \brxi^{\bri} = Y^\mu_i \xi^\nu ( \partial_\mu \bX_\nu^{\bri} - \partial_\nu \bX_\mu^{\bri}) \,.
\ea
\ee
The exterior derivatives of the one-forms $X_\mu^i$ and $\brX_\mu^{\bri}$ appearing on the right hand sides here can be interpreted as intrinsic ``torsions'' or    parts of  connections of the undoubled non-Riemannian geometry \cite{Cho:2019ofr}.
When they are vanishing, one could interpret the above equations as stating that $\xi^i$ and $\brxi^{\bri}$ depend only on the directions specified by $Y^\mu_i$ and $\brY^\mu_{\bri}$, respectively.
This will clearly be the situation in the flat non-Riemannian geometry specified by the generalised metric \eqref{mcH0}.
It would be interesting to understand the consequences of these conditions for general curved backgrounds as well as their relation to the torsion conditions that are connected to SNC~\cite{Bergshoeff:2018yvt,Bergshoeff:2019pij,Gomis:2019zyu} and TNC~\cite{Gallegos:2019icg} string theory, see also~\cite{Gallegos:2020egk}.
For now, we restrict to the flat solution~\eqref{mcH0} and turn to the full solution to the  Killing equations in this particular case.

\subsection{Most general solution for flat non-Riemannian geometry: supertranslations\label{solvingKilling}}

We present the  solution to the Killing equations for the constant  generalised metric~\eqref{mcH0}.
For this, it is natural to parametrise $(\xi^\mu,\lambda_\nu)$ as
\be
\xi^{\mu}= \left(\xi^{a}\,,\,\xi^{i}\,,\,\brxi^{\bri}\right)\,,\quad
\lambda_{\nu}= \left(\lambda_{b}\,,\,\lambda_{j}\,,\,\brlambda_{\brj}\right)\,,
\ee
and we solve for these variables that are \textit{a priori} functions of $x^{c},x^{k},\brx^{\brk}$.
The most general solution is derived in Appendix \ref{derivKilling}, and takes the form:
\begin{framed}
\be
\begin{aligned}
\xi^{a}&=w^{a}{}_{b}x^{b}+\zeta^{a}(x^{k})+\brzeta^{a}(\brx^{\brk})\,,\qquad&\qquad
\lambda_{a}&=\partial_{a}\varphi(x^{c},x^{k},\brx^{\brk})
+\zeta_{a}(x^{k})-\brzeta_{a}(\brx^{\brk})
\,,\\
\xi^{i}&=\zeta^{i}(x^{k})\,,\qquad&\qquad
\lambda_{i}&=
\partial_{i}\varphi(x^{c},x^{k},\brx^{\brk})+\rho_{i}(x^{k})\,,\\
\brxi^{\bri}&=\brzeta^{\bri}(\brx^{\brk})\,,\qquad&\qquad
\brlambda_{\bri}&=
\brpartial_{\bri}\varphi(x^{c},x^{k},\brx^{\brk})+\brrho_{\bri}(\brx^{\brk})\,,
\end{aligned}
\label{KillingSOL}
\ee
\end{framed}
where, as the arguments indicate:
\begin{enumerate}[\itshape (i)]
\item $\varphi(x^{c},x^{k},\brx^{\brk})$ is an arbitrary function of  $x^{c},x^{k},\brx^{\brk}$;
\item $\zeta^{a}(x^{k})$, $\zeta^{i}(x^{k})$, and $\rho_{i}(x^{k})$ are  arbitrary functions of
  $x^{k}$ but independent of  $x^{c}$ and $\brx^{\brk}$, so we refer to $\zeta^i(x^k)$ and $\rho_i(x^k)$ as `chiral' reparametrisations, and in analogy with BMS we refer to $\zeta^a(x^k)$ as `supertranslations';
\item $\brzeta^{a}(\brx^{\brk})$, $\brzeta^{\bri}(\brx^{\brk})$, and $\brrho_{\bri}(\brx^{\brk})$ are  arbitrary functions of  $\brx^{\brk}$, hence we refer to $\brzeta^{\bri}(\brx^\brk)$, $\brrho_{\bri}(\brx^{\brk})$ as `anti-chiral' reparametrisations and $\brzeta^{a}(\brx^{\brk})$ as supertranslations;
\item $w_{ab}=\eta_{ac}w^{c}{}_{b}=-w_{ba}$ is a skew-symmetric  constant parameter of  $\mathbf{so}(t,s)$; and
\item finally, the scalar parameter, $\varphi$, amounts to the kernel of the generalised Lie derivative: contributing to $\lambda_{\mu}$ as an exact form~$\partial_{\mu}\varphi$, it vanishes trivially in the generalised Lie derivative.

\end{enumerate}
\noindent For consistency, when $(n,\brn)=(0,0)$ we recover the usual Poincar\'e symmetry.  In the cases of $(D,0)$ or $(0,D)$,  corresponding to the two fully $\ODD$-symmetric vacua in DFT characterised  by $\cH_{AB}=\pm\cJ_{AB}$, the generalised metric Killing equations are trivially solved. Furthermore, when $n=1$ or $\brn=1$, $\rho_{i}$ or $\brrho_{\bri}$ can be absorbed into $\varphi$,  and there are no (anti-)chiral reparametrisations in the dual (tilde) directions.  Thus, we put ${\rho_{i}\equiv0}$ for ${n=1}$ and  ${\brrho_{\bri}=0}$ for ${\brn=1}$.

\subsubsection*{Commutation relation}

We write the C-bracket~\eqref{Cbracket} as
\be
\Lambda^{M}_{3}=\left[\Lambda_{1},\Lambda_{2}\right]^{M}_{\rm{C}}\,,
\ee
and note, with $\tpartial^{\mu}\equiv0$,
\be
\begin{split}
\xi_{3}^{\mu} & =
\xi_{1}^{\nu}\partial_{\nu}\xi_{2}^{\mu}-\xi_{2}^{\nu}\partial_{\nu}\xi_{1}^{\mu}\,,\\
\lambda_{3\mu}& =\xi_{1}^{\nu}\partial_{\nu}\lambda_{2\mu}-\xi_{2}^{\nu}\partial_{\nu}\lambda_{1\mu}
+\half \xi_{2}^{\nu}\partial_{\mu}\lambda_{1\nu}+\half \lambda_{2\nu}\partial_{\mu}\xi^{\nu}_{1}
-\half \xi_{1}^{\nu}\partial_{\mu}\lambda_{2\nu}-\half \lambda_{1\nu}\partial_{\mu}\xi^{\nu}_{2}\,.
\end{split}
\ee
The commutator relations between  the  most general  Killing vectors~\eqref{KillingSOL} give:
\be
\begin{aligned}
\zeta_{3}^{a}  =\zeta_{1}^{i}\partial_{i}\zeta_{2}^{a}
-w_{1}^{a}{}_{b}\zeta_{2}^{b}-\zeta_{2}^{i}\partial_{i}\zeta_{1}^{a}+w_{2}^{a}{}_{b}\zeta_{1}^{b}\,,\qquad & \qquad
\brzeta_{3}^{a}=\brzeta_{1}^{\bri}\brpartial_{\bri}\brzeta_{2}^{a}-w_{1}^{a}{}_{b}\brzeta_{2}^{b}
-\brzeta_{2}^{\bri}\brpartial_{\bri}\brzeta_{1}^{a}+w_{2}^{a}{}_{b}\brzeta_{1}^{b}\,,
\\
\zeta_{3}^{i} =\zeta_{1}^{j}\partial_{j}\zeta_{2}^{i}-\zeta_{2}^{j}\partial_{j}\zeta_{1}^{i}\,,\qquad& \qquad
\brzeta_{3}^{\bri}=\brzeta_{1}^{\brj}\brpartial_{\brj}\brzeta_{2}^{\bri}-\brzeta_{2}^{\brj}\brpartial_{\brj}\brzeta_{1}^{\bri}\,,
\\
w_{3}^{a}{}_{b}=w_{2}^{a}{}_{c}w_{1}^{c}{}_{b}-
w_{1}^{a}{}_{c}w_{2}^{c}{}_{b}\,,\qquad&
\end{aligned}
\label{COM1}
\ee
and
\be
\begin{split}
\rho_{3i}& =\zeta^{j}_{1}\partial_{j}\rho_{2i}-\half\rho_{1j}\partial_{i}\zeta^{j}_{2}-\half\zeta^{j}_{1}\partial_{i}\rho_{2j}-\zeta_{1}^{a}\partial_{i}\zeta_{2a}\,-\,(1\leftrightarrow 2)
\,,\\
\brrho_{3\bri}& =\brzeta^{\brj}_{1}\brpartial_{\brj}\brrho_{2\bri}-\half\brrho_{1\brj}\brpartial_{\bri}\brzeta^{\brj}_{2}-\half\brzeta^{\brj}_{1}\brpartial_{\bri}\brrho_{2\brj}+\brzeta_{1}^{a}\brpartial_{\bri}\brzeta_{2a}\,-\,(1\leftrightarrow 2)
\,,\\
\varphi_{3}&=\half w_{1}^{a}{}_{b}x^{b}\left(\partial_{a}\varphi_{2}-\zeta_{2a}+\brzeta_{2a}\right)+\half\!\left(\zeta_{1}^{a}+\brzeta_{1}^{a}\right)\!\partial_{a}\varphi_{2}+
\half\zeta_{1}^{i}\partial_{i}\varphi_{2}
+\half\brzeta_{1}^{\bri}\brpartial_{\bri}\varphi_{2}\,-\,(1\leftrightarrow 2)
\,.
\end{split}
\label{COM2}
\ee
The corresponding Lie algebra is spelled out below in \eqref{COMPR}.

\subsubsection*{Non-invariance of parametrisation and local Lorentz}

Although the generalised metric itself ---which we stress we wish to view as the fundamental geometric quantity--- is preserved by the Killing vectors, the adopted parametrisation is only invariant under the ordinary Lie derivative with respect to $\xi^{\mu}$ up to  existing  gauge transformations, \textit{i.e.}~the one-form gauge symmetry of the $B$-field, $\GL(n)\times\GL(\brn)$ rotations acting on the $i$ and $\bri$ indices, and the Milne-shift local symmetry~\eqref{MS}.

We illustrate this explicitly for the constant generalised metric~\eqref{mcH0}.
Given the parametrisation of the spacetime fields as in \eqref{flatH_HKB} and \eqref{flatH_XY}, one finds directly from \eqref{KillingSOL} that while $\cL_\xi H^{\mu\nu} = 0$, we have
\be
\begin{split}
\cL_\xi K_{\mu\nu} &= \deltaM K_{\mu\nu} \,,\qquad\qquad
\cL_\xi B_{\mu\nu} + 2 \partial_{[\mu} \lambda_{\nu]} = \deltaM B_{\mu\nu}\,,\\
\cL_\xi X_\mu^i &= a^i{}_j X_\mu^j \,,\qquad \qquad
\cL_\xi Y^\mu_i = - a^j{}_i Y^\mu_j + \deltaM Y^i_\mu \,, \\
\cL_\xi \bar X_\mu^{\bri} & = {\bar a}^{\bri}{}_{\brj} \brX_\mu^{\brj} \,,\qquad\qquad
\cL_\xi \brY^\mu_{\bri} = - {\bar a}^{\brj}{}_{\bri} \brY^\mu{}_{\brj} + \deltaM \brY^{\bri}_\mu \,,
\end{split}
\ee
where
\be
a^i{}_j \equiv \partial_j \zeta^i \,,\qquad
\bar a^{\bri}{}_{\brj} \equiv \bar\partial_{\brj} \bar\zeta^{\bri}\,,
\ee
give infinitesimal $\GL(n) \times \GL(\brn)$ rotations, and $\deltaM$ denotes an infinitesimal  Milne-shift  with
\be
V_{ai} = - \partial_i \zeta_a \,,\qquad
\bar V_{a\bri} = - \bar\partial_{\bri} \bar\zeta_{a} \,,\qquad
V_{ij} = \partial_i \rho_j\,,\qquad
V_{\bri \brj} = - \bar\partial_{\bri} \bar\rho_{\brj} \,.
\label{isoV}
\ee
Observe that this accounts for the persistent appearance of the one-form parameter, $\lambda_{\mu}$,  even though there is no $B$-field  apparently present.

\subsection{Dilatonic Killing equation}
\label{furtherKilling}

The minimal version of (bosonic) DFT contains in addition to the generalised metric the generalised dilaton, $d$, defined such that $e^{-2d}$ is a scalar density of weight one under generalised diffeomorphisms.
The generalised Killing equation for the generalised dilaton is thus, with ${\tpartial^{\mu}\equiv0}$,
\be
\hcL_{\Lambda}d=-\half e^{2d}\hcL_{\Lambda}\!\left(e^{-2d}\right)=
-\half e^{2d}\partial_{\mu}\left(\xi^{\mu}e^{-2d}\right)
=\xi^{\mu}\partial_{\mu}d-\half\partial_{\mu}\xi^{\mu}=0\,.
\label{Kd}
\ee
This implies for constant $d$ that the Killing vector is divergenceless
\be
\partial_{\mu}\xi^{\mu}=\partial_{a}\xi^{a}+\partial_{i}\xi^{i}+\brpartial_{\bri}\brxi^{\bri}=
\partial_{i}\zeta^{i}+\brpartial_{\bri}\brzeta^{\bri}=0\,.
\label{volume}
\ee
Since $\zeta^{i}(x^{k})$ and $\brzeta^{\bri}(\brx^{\brk})$ are distinct functions of unbarred $x^{k}$ and barred $\brx^{\brk}$, for some constant $c$ we should have
\be
\begin{aligned}
\partial_{i}\zeta^{i}=c\,,\qquad&\qquad
\brpartial_{\bri}\brzeta^{\bri}=-c\,,
\end{aligned}
\label{constantc}
\ee
and thus, they  decompose as
\be
\begin{split}
\zeta^{i}(x^{k})& =
\zeta^i_0+
\frac{c}{n} x^{i}+\frac{1}{(n-2)!}\epsilon^{ijk_{1}\cdots k_{n-2}}\partial_{j}\zeta_{k_{1}\cdots k_{n-2}}(x^{k})
\,,\\
\brzeta^{\bri}(\brx^{\brk})&=
\bar\zeta^{\bri}_0 -\frac{c}{\brn} \brx^{\bri}+\frac{1}{(\brn-2)!}\brepsilon^{\bri\brj\brk_{1}\cdots \brk_{\brn-2}}\brpartial_{\brj}\zeta_{\brk_{1}\cdots\brk_{n-2}}(\brx^{\brk})\,,
\end{split}
\label{volumeSOL}
\ee
exhibiting a constant shift, scaling symmetry and a volume-preserving diffeomorphism.
Note that when $n=1$ or $\brn=1$ the final terms are absent.
We may also note that in the commutators \eqref{COM1} we get
\be
\partial_{i}\zeta^{i}_{3}=0\,,\qquad
\brpartial_{\bri}\brzeta_{3}^{\bri}=0\,,
\ee
hence the scaling constant $c$ in \eqref{constantc} becomes trivial after commutations.
Although in general we have to take these extra conditions into account, in certain situations ---such as at the classical level on the string worldsheet, which we study below--- the dilaton does not appear and we will be able to make use of the full set of the isometries of \eqref{KillingSOL}.

\subsection{DFT Killing spinor equation and supersupersymmetry}
\label{KillingSpinor}

In this subsection, we discuss the supersymmetry transformations preserving the constant non-Riemannian background \eqref{mcH0}.
The starting point is to consider the coset parametrised by the generalised metric, which featured the doubled Lorentz group, $\mathbf{O}(t+n,s+n)\times\mathbf{O}(s+\brn,t+\brn)$,\footnote{Note the signature convention implicit here, which sets the `total' temporal and spatial dimensions, as for $\ODD$,  to be the same as ${D=t+s+n+\brn}$. It treats the doubled vielbeins $V_{Ap}$ and $\brV_{A\brp}$ in a `fair' manner and removes minus signs in many formulas, especially in the full order (\textit{i.e.~}quartic) supersymmetric completion~\cite{Jeon:2012hp}.} where $t,s,n,\bar n$ are fixed numbers, and for which we define spinors of each factor in the usual manner.
The minimal spinor depends on the values of $t$ and $s$.
The value of $(t-s)$ mod 8 determines what reality conditions can be imposed, while for $t+s$ even, we can have Weyl spinors of the $(t+s+2n)$- and $(t+s+2\brn)$-dimensional spin groups.

We will apply results from the known formulations of type II  supersymmetric double field theory, in particular \cite{Jeon:2012hp} which, having the Minkowskian  spin group $\Spin(1,9)\times\Spin(9,1)$, of course covers the Riemannian type II backgrounds  of  $(t,s,n,\brn) = (1,D-1,0,0)$ with $D=10$, and  is also immediately applicable to   $(t,s,n,\brn) = (0,D-2,1,1)$  and thus to non-relativistic strings. Furthermore, by relaxing the Majorana condition therein and adopting not the  Dirac but  charge conjugation of spinors, the constructed supersymmetric double field theory is readily generalised  to an arbitrary signature of the spin group with ${n=\brn}$.    The invariant metrics of the doubled Lorentz group are then
\be
\ba{ll}
\eta_{pq}=
\begin{pmatrix}
\eta_{ab} & 0 & 0 \\
0 & - \delta_{ij} & 0 \\
0 & 0 & + \delta_{ij}
\end{pmatrix}
\,,\quad&\quad
\eta_{ab}=\mbox{diag}(\underbrace{-\,-\,\cdots\, -\,-}_{t}\,\underbrace{+\,+\,\cdots \,+\,+}_{s})\,,\\

\breta_{\brp\brq}=
\begin{pmatrix}
\breta_{\bra\brb} & 0 & 0 \\
0 & +\delta_{\bri\brj} & 0 \\
0 & 0 & - \delta_{\bri\brj}
\end{pmatrix}
\,,\quad&\quad
\breta_{\bra\brb}=\mbox{diag}(\underbrace{+\,+\,\cdots\, +\,+}_{t}\,\underbrace{-\,-\,\cdots \,-\,-}_{s})\,.
\ea
\label{etas}
\ee
Note that the flat indices $p$ and $\brp$ have ranges $p=1,2,\dots,t+s+2n$ and  $\brp =1,2,\dots,t+s+2\brn$. Furthermore, we have introduced new indices $a,\bra =1,\dots, s+t$ to run over the common $(s+t)$-dimensional part (Riemannian), and decompose $p=(a,i,n+j)$ and $\brp=(\bra,\bri,\brn+\brj)$.

The bosonic degrees of freedom can then be recast in terms of a pair of DFT-vielbeins $V_{A p}$ and $\brV_{A \brp}$, which square to the orthogonal projectors whose existence is implied by the  defining property of the generalised metric~\eqref{defcH}.  Namely:
\be
P_{AB} = \tfrac{1}{2} ( \cJ_{AB} + \cH_{AB} ) = V_{Ap} V_{Bq} \eta^{pq} \,,\qquad
\brP_{AB} = \tfrac{1}{2} ( \cJ_{AB} - \cH_{AB} ) = \brV_{A\brp} \brV_{B\brq} \breta^{\brp\brq} \,.
\label{PbrPVbrV}
\ee
Note this means that $\cH_{AB} = V_{Ap} V_{Bq} \eta^{pq} -  \brV_{A\brp} \brV_{B\brq} \breta^{\brp\brq}$ and
$\cJ_{AB} = V_{Ap} V_{Bq} \eta^{pq} +\brV_{A\brp} \brV_{B\brq} \breta^{\brp\brq}$ are simultaneously  diagonalised by $(V_{A}{}^p, \brV_{A}{}^{\brp})$ with  the expected signatures, $(\eta,-\breta)$ and $(\eta,\breta)$ respectively. From identities of arbitrary variations like   $\delta V_{Ap}=V_{A}{}^{q}V_{B[q}\delta V^{B}{}_{p]}+\brP_{A}{}^{B}\delta V_{Bp}$ and $\delta \brV_{A\brp}=\brV_{A}{}^{\brq}\brV_{B[\brq}\delta \brV^{B}{}_{\brp]}+P_{A}{}^{B}\delta\brV_{B\brp}$~\cite{Jeon:2011sq}, the generalised metric Killing equation~\eqref{Killing} implies that the generalised Lie derivatives of the vielbeins  vanish  only up to certain  infinitesimal  local  Lorentz rotations:
\be
\begin{aligned}
\hcL_{\Lambda}V_{Ap}+(V_{B[p}\hcL_{\Lambda}V^{B}{}_{q]})V_{A}{}^{q}=+\half V^{B}{}_{p}\hcL_{\Lambda}\cH_{AB}\equiv0\,,&\\
\hcL_{\Lambda}\brV_{A\brp}+(\brV_{B[\brp}\hcL_{\Lambda}\brV^{B}{}_{\brq]})\brV_{A}{}^{\brq}=-\half \brV^{B}{}_{p}\hcL_{\Lambda}\cH_{AB}\equiv0\,.&
\end{aligned}
\label{KVV}
\ee
An explicit parametrisation of these vielbeins is~\cite{Morand:2017fnv}:
\be
\begin{aligned}
&V_{A p} = \tfrac{1}{\sqrt{2}} \begin{pmatrix} h^\mu{}_p \\ k_{\mu p} + B_{\mu\nu} h^\nu{}_p \end{pmatrix}\,,\qquad
&\brV_{A \brp} = \tfrac{1}{\sqrt{2}} \begin{pmatrix} \brh^\mu{}_{\brp} \\ \brk_{\mu\brp} + B_{\mu\nu} \brh^\nu{}_{\brp} \end{pmatrix}\,,
\end{aligned}
\ee
where in order to reproduce the generic $(n,\brn)$ generalised metric \eqref{cHFINAL} we let
\be
\begin{gathered}
h^\mu{}_p \equiv \left( h^\mu{}_a , Y^\mu_i, Y^\mu_i \right) \,,\quad
\brh^\mu{}_{\brp} \equiv\left( \brh^\mu{}_{\bra} , \brY^\mu_{\bri}, \brY^\mu_{\bri}\right) \,,
\\
k_\mu{}^p \equiv \left( k_\mu{}^a , X_\mu^i, X_\mu^i \right) \,,\quad
\brk_{\mu}{}^{\brp} \equiv \left( \brk_\mu{}^{\bra} , \brX_\mu^{\bri}, \brX_\mu^{\bri} \right) \,.
\end{gathered}
\ee
Here we have  $H^{\mu\nu} = h^{\mu}{}_a h^{\nu}{}_{b} \eta^{ab}=-\brh^{\mu}{}_{\bra} \brh^{\nu}{}_{\brb} \breta^{\bra\brb}$  and $K_{\mu\nu} = k_{\mu}{}^a k_{\nu}{}^{b} \eta_{ab}=-\brk_{\mu}{}^{\bra} \brk_{\nu}{}^{\brb} \breta_{\bra\brb}$  with $\eta_{ab}$, $\breta_{\bra \brb}$ given in \eqref{etas}. Note also $k_{\mu}{}^{a}h^{\mu}{}_{b}=\delta^{a}{}_{b}$ and $
\brk_{\mu}{}^{\bra}\brh^{\mu}{}_{\brb}=\delta^{\bra}{}_{\brb}$, while $k_\mu{}^a h^\nu{}_a + X_\mu^i Y^\nu_i + \bX_\mu^{\bri} \bY^\nu_{\bri} = \brk_\mu{}^{\bra} \brh^\nu{}_{\bra} + X_\mu^i Y^\nu_i + \bX_\mu^{\bri} \bY^\nu_{\bri} = \delta_\mu{}^\nu$.

The fermions in type II supersymmetric double field theory consist of pairs of gravitinos and dilatinos,  $\left\{\psi_{\brp}^{\alpha},\psi_{p}^{\prime\bralpha},\rho^{\alpha},\rho^{\prime\bralpha}\right\}$, which are all Weyl spinors with appropriate chiralities.\footnote{Since $\Spin(t+n,s+n)$ and $\Spin(s+\brn,t+\brn)$ are independent, the comparison between the corresponding chiralities is meaningless. It is the paired vielbeins, $V_{Ap},V_{A\brp}$, that distinguish type IIA and IIB~ \cite{Jeon:2012hp} (see footnote~\ref{footIIAB}).}  As the indices indicate, they are spinors for only one of the doubled spin group and singlets for the other. Furthermore, the gravitinos carry an `opposite' vector index.  In a constant background (with vanishing fermions) the supersymmetry transformations of the fermions are~\cite{Jeon:2012hp}:
\be
\begin{aligned}
\delta_{\varepsilon} \psi^{\alpha}_{\brp} & = \brV^A{}_{\brp} \partial_A \varepsilon^{\alpha} \,, \qquad&\qquad \delta_{\varepsilon} \rho^{\alpha}&  =  - (\gamma^{p})^{\alpha}{}_{\beta}  V^A{}_{p}\partial_A \varepsilon^{\beta}\,,\\
\delta_{\varepsilon} \psi_{p}^{\prime\bralpha} & = V^A{}_{p} \partial_A \varepsilon^{\prime\bralpha} \,,\qquad&\qquad
\delta_{\varepsilon} \rho^{\prime\bralpha} &= -(\brgamma^{\brp})^{\bralpha}{}_{\brbeta}\brV^A{}_{\brp} \partial_A \varepsilon^{\prime\brbeta} \,.
\end{aligned}
\label{SUSYvars}
\ee
The two sets of gamma matrices, $\gamma^p= (\gamma^a, \gamma^{i}, \gamma^{n+j})$,  $\brgamma^{\brp}= (\brgamma^{\bra}, \brgamma^{\bri}, \brgamma^{\brn+\brj})$ realise the Clifford algebras of $\Spin(t+n,s+n)$ and $\Spin(s+\brn,t+\brn)$ respectively, so $\{ \gamma^p, \gamma^q \} = 2 \eta^{pq}$ and $\{ \brgamma^{\brp}, \brgamma^{\brq} \} = 2 \breta^{\brp\brq}$ with the flat metrics given in \eqref{etas}.   With  the choice of the section, ${\tpartial^{\mu}\equiv0}$, we get $V^{A}{}_{p}\partial_{A}=\tfrac{1}{\sqrt{2}} h^{\mu}{}_{p}\partial_{\mu}$ and $\brV^{A}{}_{\brp}\partial_{A}=\tfrac{1}{\sqrt{2}} \brh^{\mu}{}_{\brp}\partial_{\mu}$.

Now, specifically for the constant flat  $(n,\brn)$ generalised  metric~\eqref{mcH0} with the coordinates, $x^{\mu}=(x^{a},x^{i},\brx^{\bri})$~\eqref{flatcoords} and $B_{\mu\nu}=0$, we can take
\be
\begin{gathered}
h^\mu{}_a = \delta^\mu_a\,,\quad
\brh^\mu{}_{\bra} = \delta^\mu_{\bra}\,,\quad
Y^\mu_i = \delta^\mu_i\,,\quad
\brY^\mu_{\bri} = \delta^\mu_{\bri}\,\\
k_\mu{}^a = \delta_\mu^a\,,\quad
\brk_\mu{}^{\bra} = \delta_\mu^{\bra}\,,\quad
X_\mu^i = \delta_\mu^i\,,\quad
\brX_\mu^{\bri} = \delta_\mu^{\bri}\,,
\end{gathered}
\label{flatV}
\ee
and identify the flat indices $a\equiv\bra$ via a compensating $\mathrm{O}(t,s)$ Lorentz rotation.\footnote{Depending on the determinant of this transformation, or the sign of $\det(k_{\mu}{}^{a}\brh^{\mu}{}_{\bra})=\pm 1$,   we may distinguish    two distinct classes of backgrounds, as the compensating Lorentz rotation is generically  $\Pin$  rather than $\Spin$. This generalises the usual distinction of  type IIA and  IIB of the Riemannian case to the non-Riemannian case.
\label{footIIAB}}
Setting the variations \eqref{SUSYvars} to zero we find from the variations of the gravitini the conditions that
\be
\ba{ll}
\partial_a \varepsilon = \bar\partial_{\bri} \varepsilon=0\,,\qquad&\qquad
\partial_a \varepsilon^\prime = \partial_i \varepsilon^\prime =0\,.
\ea
\ee
Then, from the variation of the dilatini we have the requirement
\be
\ba{ll}
\gamma_{+}^{i}\partial_{i}\varepsilon=0\,,\qquad&\qquad
\brgamma_{+}^{\bri}\brpartial_{\bri}\varepsilon^{\prime}=0\,,
\ea
\label{dilatiniK}
\ee
where we set $\gamma_{+}^{i}=\gamma^{i}+\gamma^{n+i}$ and $\brgamma_{+}^{\bri}=\brgamma^{\bri}+\brgamma^{\brn+\bri}$ which satisfy
\be
\ba{llll}
\gamma_{+}^{i}\gamma_{+}^{j}+\gamma_{+}^{j}\gamma_{+}^{i}=0\,,
\qquad&\left(\gamma_{+}^{i}\partial_{i}\right)^{2}=0\,,\qquad&\qquad
\brgamma_{+}^{\bri}\brgamma_{+}^{\brj}+\brgamma_{+}^{\brj}\brgamma_{+}^{\bri}=0\,,
\qquad&\left(\brgamma_{+}^{\bri}\brpartial_{\bri}\right)^{2}=0\,.
\ea
\ee
Thus, in the poly-form representation of spinors,  $\gamma_{+}^{i}$ and $\gamma_{+}^{i}\partial_{i}$ correspond to the coordinate basis of one-forms $\rd x^{i}$ and the exterior derivative $\rd=\rd x^{i}\partial_{i}$ (similarly $\brgamma_{+}^{\bri}\rightarrow \rd\brx^{\bri}$).  Then  from the Poincar\'{e} lemma, the supersymmetric condition of \eqref{dilatiniK} implies that the supersymmetric parameters are `closed' forms, and locally `exact' except the lowest zero-form which should be simply constant.
In conclusion, the most general (at least locally\footnote{Global solutions with nontrival cohomology would be of interest but are beyond the scope of the present work.}) solution to the  Killing spinor equations~\eqref{SUSYvars} is
\begin{framed}
\be
\ba{ll}
\varepsilon = ( \gamma^{j} + \gamma^{n+j})\partial_{j}\chi(x^k)+\epsilon \,,\qquad&\qquad
\varepsilon^\prime = ( \brgamma^{\brj} + \brgamma^{\brn+\brj} )\brpartial_{\brj}\brchi(\brx^{\brk})+\epsilon^\prime \,,
\ea
\label{GSKS}
\ee
\end{framed}
\noindent where $\chi(x^{k})$ and $\brchi(\brx^{\brk})$ are spinors with the opposite chiralities to $\varepsilon$ and $\varepsilon^{\prime}$ respectively, and arbitrarily (or supertranslationally) depend on  $x^{k}$ and $\brx^{\brk}$. Furthermore, $\epsilon $ and $\epsilon^{\prime}$ are constant spinors which survive in the Riemannian $(0,0)$ case.

In the particular case of $(n,\brn)=(1,1)$ with $D=10$, the derivatives in \eqref{GSKS} are redundant.
The indices $i$ and $\bri$ cover only a single value, and so if we write\footnote{Given the form of the flat metrics in \eqref{etas}, here we have $-(\gamma^0)^2 =  (\gamma^1)^2 = 1$, $\gamma^0 \gamma^1 = - \gamma^1 \gamma^0$, while $(\brgamma^{\bar 0})^2 =  - (\brgamma^{\bar1})^2 = 1$, $\brgamma^{\bar 0} \brgamma^{\bar 1} = - \brgamma^{\bar 1} \brgamma^{\bar 0}$.
Let us also note that with the assumed identification of $a\equiv\bra$,  we set $\gamma_{a}\equiv\delta_{ab}\gamma^{b}$ and $\brgamma_{a}\equiv\delta_{ab}\brgamma^{b}$. (This introduces a minus sign in certain expressions owing to the fact that the barred flat metric has components $\breta_{\bar a \bar b} = -\delta_{\bar a \bar b} = - \delta_{ab}\,$.)}
\be
\gamma^p \equiv (\gamma^a, \gamma^0, \gamma^1 ) \,,\qquad\quad
\brgamma^{\brp} \equiv ( \brgamma^a, \brgamma^{\bar 0}, \brgamma^{\bar 1} )\,,
\ee
then the general solution can be written
\be
\ba{ll}
\varepsilon = ( \gamma^{0} + \gamma^{1}) \chi(y)+\epsilon\,, \quad&\quad
\varepsilon^\prime = ( \brgamma^{\bar{0}} + \brgamma^{\bar{1}} )\chi^\prime(\bry) +\epsilon^{\prime}\,.
\ea
\label{SpinorSOL}
\ee
where $y=x^{1}$ and $\bry=\brx^{1}$ denote the coordinates corresponding to  the $n=1$ and $\brn =1$ non-Riemannian directions respectively.

The commutator of two SUSY transformations generates  ---among other local symmetries \cite{Jeon:2012hp}--- a DFT diffeomorphism with parameter
\be
\Lambda^A =
i \bar \varepsilon_2^\prime \brV^{A}{}_{\brp} \brgamma^{\brp} \varepsilon_1^\prime
+ i \bar \varepsilon_2 V^{A}{}_{p} \gamma^{p} \varepsilon_1 \,,
\label{SUSYCOMM}
\ee
where $\bar{\varepsilon}=\varepsilon^{T}C$ and  $\bar{\varepsilon}^{\prime}=\varepsilon^{\prime T}\brC$ are charge conjugations.\footnote{The charge conjugation matrices are all symmetric, $C=C^{T}$, $\brC=\brC^{T}$,  and satisfy    $(\gamma^{p})^{T}=C\gamma^{p}C^{-1}$, $(\brgamma^{\brp})^{T}=\brC\brgamma^{\brp}\brC^{-1}$, hence $\overline{ \gamma^p \varepsilon } = \bar\varepsilon \gamma^p$, $\overline{ \brgamma^{\brp} \varepsilon^\prime } = \bar\varepsilon^\prime \brgamma^{\brp}$. For further  details, see   Appendix  of \cite{Jeon:2012kd}. }
This is then easily computed for \eqref{SpinorSOL}.
For the  different spinor bilinears that appear, let us set
\be
c^a \equiv \tfrac{i}{\sqrt{2}} \bar\epsilon_2  \gamma^a  \epsilon_1\,,\quad
\bar c^a \equiv \tfrac{i}{\sqrt{2}} \bar\epsilon^\prime_2  \brgamma^a  \epsilon^\prime_1\,,\quad
c^y  \equiv \tfrac{i}{\sqrt{2}} \bar\epsilon_2 ( \gamma^0+\gamma^1) \epsilon_1 \,,\quad
\bar c^{\bar y} \equiv \tfrac{i}{\sqrt{2}} \bar\epsilon_2^\prime ( \brgamma^{\bar{0}}+\brgamma^{\bar{1}}) \epsilon_1^\prime\,,
\ee
\be
\begin{split}
\zeta^a(y)& \equiv \tfrac{i}{\sqrt{2}} \left[  \bar \epsilon_2  \gamma^a (\gamma^0+\gamma^1)  \chi_1(y)
- \bar \chi_2 (y) \gamma^a (\gamma^0+\gamma^1) \epsilon_1 \right]\,,\\
\bar\zeta^a(\bar y) &\equiv \tfrac{i}{\sqrt{2}}  \left[ \bar \epsilon^\prime_2  \brgamma^a (\brgamma^{\bar{0}}+\brgamma^{\bar{1}})  \chi^\prime_1(\bry)
- \bar\chi_2^\prime(\bar y) \brgamma^a (\brgamma^{\bar{0}} + \brgamma^{\bar{1}} ) \epsilon_1^\prime\right]\,.
\end{split}
\ee
Then on inserting \eqref{SpinorSOL} into \eqref{SUSYCOMM}, writing as before $\Lambda^A = ( \lambda_\mu, \xi^\nu)$, one finds firstly that
\be
\xi^a = \epsilon^a + \bar\epsilon^a  + \zeta^a(y) + \bar\zeta^a(\bar y) \,,\qquad
\lambda_a = \epsilon_a - \bar\epsilon_a + \zeta_a(y) - \bar\zeta_a(\bar y) \,,
\ee
which is exactly in agreement with the form of \eqref{KillingSOL} (on absorbing the constant parts $\epsilon^a$, $\bar\epsilon^a$ into $\zeta^a, \bar\zeta^a$).
We further have
\be
\xi^y = c^y \,,\qquad \xi^{\bar y} =\bar c^{\bar y}\,,
\ee
which are constant shifts in the $y$ and $\bar y$ directions rather than shifts by arbitrary functions of $y$ and $\bar y$, however this is to be expected as we have made use of the Killing spinor equations for not only the gravitinos but also the dilatinos, and we already know that including the dilaton as in section \ref{furtherKilling} restricts the solutions as in \eqref{volumeSOL} (here only the constant shift part is generated).
Finally, we have
\begin{align}
\lambda_y
&=
\tfrac{-i}{\sqrt{2}} \bar\epsilon_2 (\gamma^0-\gamma^1)  \epsilon_1
\\\nonumber
&{}\qquad
+ \sqrt{2} i \Big( \bar \epsilon_2  (1-\gamma^0\gamma^1) \chi_1(y)
+  \bar \chi_2 (y)  (1+\gamma^0\gamma^1)\epsilon_1
+2  \bar \chi_2(y)  (\gamma^0+\gamma^1) \chi_1(y) \Big),
\\
\lambda_{\bar y}
&=
\tfrac{i}{\sqrt{2}} \bar\epsilon^\prime_2 (\brgamma^{\bar{0}}-\brgamma^{\bar{1}})  \epsilon^\prime_1
\\\nonumber
&{}\qquad
+ \sqrt{2} i \Big( \bar \epsilon^\prime_2  (1+\brgamma^{\bar{0}}\brgamma^{\bar{1}}) \chi^\prime_1 (\bar y)
+  \bar \chi^\prime_2 (\bar y) (1-\brgamma^{\bar{0}}\brgamma^{\bar{1}}) \epsilon^\prime_1
+2 \bar \chi^\prime_2 (\bar y) (\brgamma^{\bar{0}}+\brgamma^{\bar{1}}) \chi^\prime_1(\bar y)\Big) ,
\end{align}
which  again agrees with   \eqref{KillingSOL}.

\section{Sigma models and worldsheet Noether charges}
\label{StringSection}

We now turn our attention to the description of strings whose target spacetime is described by a generalised metric corresponding to a non-Riemannian geometry.
This has already been considered in general in \cite{Morand:2017fnv,Park:2020ixf} for the bosonic string, and in more specific supersymmetric cases in \cite{Park:2016sbw, Blair:2019qwi}.
Here we review the general bosonic case in order to extract expressions for the conserved worldsheet Noether  charges induced by the doubled target  spacetime  Killing isometries.
We also make some remarks regarding general features of these models, and argue that they generalise the worldsheet action of stringy Newton--Cartan (SNC) non-relativistic strings to arbitrary non-Riemannian backgrounds.

\subsection{Sigma model in general background}
\label{SigmaModelAction}

The doubled  $\ODD$-symmetric sigma model string action that we use is~\cite{Hull:2004in, Hull:2006va,Lee:2013hma}
\be
\begin{split}
S_{\text{{DWS}}} = - \frac{T}{2} \int d^2\sigma~ \half \sqrt{-\gammaws} \gammaws^{\alpha \beta} \cH_{AB} \rD_\alpha x^A \rD_\beta x^B + \epsilon^{\alpha \beta} \cJ_{AB} \rD_\alpha x^A \mathcal{A}_\beta^B \,.
\end{split}
\label{doubleWS}
\ee
Here, $\gammaws_{\alpha\beta}$ is the worldsheet metric, and $\epsilon^{\alpha\beta}$ is the worldsheet alternating symbol with $\epsilon^{01} = 1$ (a useful relation is $\epsilon^{\alpha \gamma}\epsilon^{\beta\delta} h_{\gamma\delta} = (\det h) h^{\alpha \beta}$).
This is an action for a set $x^A = (\tilde x_\mu, x^\nu)$ of $2D$ worldsheet scalars that we treat as doubled target space coordinates.
Due to the section condition~\eqref{sectioncondition}, the background generalised metric is independent of half the doubled coordinates $x^A$.
These ``unphysical'' coordinates therefore appear in the action \eqref{doubleWS} solely through their worldsheet derivatives, and in fact come with a shift symmetry.
As was argued in~\cite{Park:2013mpa}\footnote{In \cite{Park:2013mpa}, the section condition~\eqref{sectioncondition} was interpreted as gauging the doubled coordinates by imposing an  equivalence relation: $x^{A}\sim x^{A}+\Delta^{A}$ where $\Delta^{A}\partial_{A}=0$.}  we can \textit{gauge} this symmetry, which in the worldsheet action~\cite{Hull:2004in, Hull:2006va, Lee:2013hma} allows us to eliminate half of the doubled coordinates from the action.
To this end, we introduce a worldsheet one-form $\cA^A$, which we take to satisfy a gauging constraint $\cA^A\, \partial_A = 0$.
This is present in the action~\eqref{doubleWS} through the combination $\rD x^A = \rd x^A - \cA^A$ (note $\rD x^A \equiv \rd\sigma^\alpha\rD_\alpha x^A $), and $\cA^A$ is required to transform such that $\rD x^A$ is invariant under the shift symmetry of the unphysical coordinates.
To realise isometries of the spacetime as Noether symmetries of the worldsheet, $\cA^A$ is also required to transform under generalised diffeomorphisms, along with $x^A$~\cite{Lee:2013hma}:
\be
\delta_\Lambda x^{A}=\Lambda^{A}\,,\qquad
\delta_\Lambda \cA^{A}=\rmD x^{B}\partial^{A}\Lambda_{B}\,.
\label{diffeo}
\ee
As a result, we have $\delta_\Lambda(\rd x^{A})=\rd x^{B}\partial_{B}\Lambda^{A}$, $\delta_\Lambda(\rmD x^{A})=\rmD x^{B}(\partial_{B}\Lambda^{A}-\partial^{A}\Lambda_{B})$, and hence the transformation of the action \eqref{doubleWS} is, with $\delta_{\Lambda}\cH_{AB}=\Lambda^{C}\partial_{C}\cH_{AB}$ (as $\cH_{AB}$ is regarded on the worldsheet merely as a set of functions depending on the worldsheet coordinates),
\be
\delta_\Lambda S_{\text{DWS}}
= - \frac{T}{2} \int d^2\sigma \half \sqrt{-\gammaws} \gammaws^{\alpha \beta} ( \hcL_\Lambda \cH_{AB} ) \rD_\alpha x^A \rD_\beta x^B
+ \partial_\alpha ( \epsilon^{\alpha \beta} \Lambda_A \partial_\beta x^A ) \,.
\ee
In particular, if $\Lambda$ is a generalised Killing vector, we get a symmetry of the action up to total derivatives, and Noether's theorem implies the following current is on-shell conserved:
\be
\mathcal{J}_\Lambda^\alpha
= \frac{T}{2} \Lambda^A \left(
\epsilon^{\alpha \beta} \rD_\beta x_A
-
\sqrt{-\gammaws} \gammaws^{\alpha \beta} \cH_{AB} \rD_\beta x^B
\right)\,.
\label{stringJ}
\ee
Coupled to the (parametrisation independent) generalised metric, the above  sigma model~\eqref{doubleWS} can describe a number of different target space geometries.
To interpret these in $D$-dimensional terms, we solve the section condition as ${\tilde \partial^\mu \equiv 0}$ acting on fields, which implies  $\cA^A = ( \tilde \cA_{ \mu},0)$,  $\delta_{\Lambda}(\rd x^{\mu})=\rd x^{\nu}\partial_{\nu}\xi^{\mu}$,  $\delta_{\Lambda}(\rD \tx_{\mu})=-\partial_{\mu}\xi^{\nu}\rD\tx_{\nu}-2\partial_{[\mu}\lambda_{\nu]}\rd x^{\nu}$, and spontaneously breaks the formal $\ODD$ invariance of the action.
Integrating out the non-zero components $\tilde \cA_{ \mu}$ then has the effect of ``undoubling'' the string action.
Let us describe this explicitly for Riemannian and non-Riemannian cases.

\subsubsection*{Riemannian geometry}

Suppose the generalised metric is described by the $(0,0)$ parametrisation \eqref{Hnormal}, describing a Riemannian metric and a $B$-field.
In this case, the worldsheet gauge field $\cA_\alpha^A$ appears quadratically in the action and can be completely integrated out, leading to the constraint,
\be
\rD_\alpha \tilde x_\mu = - \frac{1}{\sqrt{-\gammaws}} h_{\alpha \beta} \epsilon^{\beta \gamma} g_{\mu\nu} \partial_\gamma x^\nu + B_{\mu\nu} \partial_\alpha x^\nu \,,
\label{dualityrelation}
\ee
and (after eliminating a total derivative\footnote{Possibly through a gauge fixing~\cite{Park:2020ixf}.}) the usual (Riemannian) string action:
\be
S_{\text{DWS}} \stackrel{\text{eliminate $\tilde x$}}{\xrightarrow{\hspace*{1.5cm}}} S =\frac{T}{2} \int d^2\sigma\, - \sqrt{-\gammaws} \gammaws^{\alpha \beta} g_{\mu\nu} \partial_\alpha x^\mu \partial_\beta x^\nu
+ \epsilon^{\alpha \beta} B_{\mu\nu} \partial_\alpha x^\mu \partial_\beta x^\nu\,.
\ee

\subsubsection*{Stringy Newton--Cartan}

If we consider the generalised metric \eqref{HSNC} describing stringy Newton--Cartan, then the block $\mathcal{H}^{\mu\nu}$ is non-invertible.
As a result, one cannot integrate out all components of $\cA^A$ from the sigma model action.
Instead, we can use the completeness relation \eqref{SNCcomplete} to write $\tilde\cA_{ \mu}$ as a sum of a piece orthogonal to $v^\mu{}_M$ and a piece proportional to $v^\mu{}_M$.
We can integrate out the former. (The calculation is essentially identical to the more general case below, for which we will present more details.)
The result is the Polyakov action for SNC \cite{Bergshoeff:2018yvt, Harmark:2019upf}:
\be
\begin{split}
S_{\text{DWS}} \stackrel{\text{eliminate $\tilde x$}}{\xrightarrow{\hspace*{1.5cm}}}
S_{\text{SNC}} = \frac{T}{2} \int d^2 \sigma& \, - \sqrt{-\gammaws} \gammaws^{\alpha \beta} H^\perp_{\mu\nu} \partial_\alpha x^\mu \partial_\beta x^\nu + \epsilon^{\alpha \beta} \bar B_{\mu\nu} \partial_\alpha x^\mu \partial_\beta x^\nu
\\& \quad
+ \beta_{\alpha}{}^{\lA}  \big(- \sqrt{-\gammaws} \gammaws^{\alpha \beta} \epsilon_{\lA\lB} -\epsilon^{\alpha \beta} \eta_{\lA\lB} \big) \tau_\mu{}^{\lB} \partial_\beta x^\mu \,,
\end{split}
\ee
where the surviving components of $\cA_\alpha^A$ appear in the combination
\be
\beta_\alpha{}^{\lA} \equiv v^{\mu \lA} ( \rD_{\alpha} \tilde x_\mu - \bar B_{\mu\nu} \partial_\alpha x^\nu ) \,,
\ee
which we can treat as an independent worldsheet field.
We can rewrite this after splitting the longitudinal coordinates as ${\lA = (+,-)}$ with ${\epsilon_{+-} = -1}$, ${\eta_{+-} =1}$, ${\eta_{++} = \eta_{--} = 0}$.
Then
\be
\begin{split}
S_{\text{SNC}} = \frac{T}{2} \int d^2 \sigma& \, - \sqrt{-\gammaws} \gammaws^{\alpha \beta} H^\perp_{\mu\nu} \partial_\alpha x^\mu \partial_\beta x^\nu + \epsilon^{\alpha \beta} \bar B_{\mu\nu} \partial_\alpha x^\mu \partial_\beta x^\nu
\\& \quad
+ \beta_{\alpha}{}^-  ( - \sqrt{-\gammaws} \gammaws^{\alpha \beta}  -\epsilon^{\alpha \beta}  ) \tau_\mu{}^{+} \partial_\beta x^\mu
- \beta_{\alpha}{}^+  ( - \sqrt{-\gammaws} \gammaws^{\alpha \beta}  +\epsilon^{\alpha \beta}  ) \tau_\mu{}^{-} \partial_\beta x^\mu \,,
\end{split}
\ee
which exhibits a split into chiral and anti-chiral directions.
At this point, we may note that there is a freedom to perform field redefinitions of the $\beta$ fields, along the lines of those discussed in \cite{Harmark:2019upf}.

Another route to stringy non-relativistic geometries involves T-dualising along a null isometry direction, leading to strings in torsional Newton--Cartan geometry \cite{Harmark:2017rpg,Harmark:2018cdl}.
Although this would naively be singular using the usual Buscher prescription, such T-duality transformations act on the generalised metric simply as a permutation of its components, and lead in the case of a transformation along a null isometry direction to a generalised metric for which $\cH^{\mu\nu}$ is degenerate. A dictionary between SNC and TNC parametrisations was worked out in \cite{Harmark:2019upf}, and the generalised metric parametrisations described in detail in  \cite{Berman:2019izh,Blair:2019qwi}.

\subsubsection*{Non-Riemannian geometries}

For the doubled sigma model~\eqref{doubleWS} corresponding to a string in a general $(n,\brn)$ non-Riemannian geometry, we wish to integrate out the dual coordinates using the equation of motion for $\tilde\cA_{ \mu}$.
To this end, it is convenient to first define the combination
\be
P_{\alpha \mu} \equiv \partial_\alpha \tilde x_\mu - \tilde\cA_{\alpha \mu} - B_{\mu\nu} \partial_\alpha x^\nu \,.
\label{defP}
\ee
This turns out to be (on-shell) directly related to the momentum current of the undoubled sigma model, hence the notation $P_{\alpha \mu}$.
For an $(n,\brn)$ parametrisation, we expand this using the completeness relation \eqref{COMP} and thus define $P_{\alpha \mu} = \Pi_{\alpha \mu} + X_\mu^i \beta_{\alpha i} - \brX_\mu^{\bri} \bar \beta_{\alpha \bri}$ with
\be
\Pi_{\alpha \mu} \equiv K_{\mu\rho} H^{\rho \nu} P_{\alpha \nu} \,,\quad
\beta_{\alpha i} \equiv Y^\mu_i P_{\alpha \mu}\,,\quad
\bar\beta_{\alpha \bri} \equiv - \bar Y^\mu_{\bri} P_{\alpha \mu} \,,
\label{defbetas}
\ee
where the minus sign in the final definition is for convenience.
The doubled sigma model action becomes:
\begin{align}
\label{nnaction}
S_{\text{DWS}} & =\frac{T}{2} \int d^2\sigma\, - \sqrt{-\gammaws} \gammaws^{\alpha \beta} K_{\mu\nu} \partial_\alpha x^\mu \partial_\beta x^\nu
+ \epsilon^{\alpha \beta} B_{\mu\nu} \partial_\alpha x^\mu \partial_\beta x^\nu
\\\nonumber & \qquad\qquad
+ \beta_{\alpha i} X_{\mu}^i\left ( - \sqrt{-\gammaws} \gammaws^{\alpha \beta} \partial_\beta x^\mu - \epsilon^{\alpha \beta} \partial_\beta x^\mu\right)
+ \bar \beta_{\alpha \bri} \bar X_\mu^{\bri}\left( -\sqrt{-\gammaws} \gammaws^{\alpha \beta} \partial_\beta x^\mu + \epsilon^{\alpha \beta} \partial_\beta x^\mu\right)
\\\nonumber & \qquad\qquad
- \half \sqrt{-\gammaws} \gammaws^{\alpha \beta} H^{\mu\nu}
\left(
\Pi_{\alpha \mu} + \frac{1}{\sqrt{-\gammaws}} \gammaws_{\alpha \gamma}\epsilon^{\gamma\delta} K_{\mu\rho} \partial_\delta x^\rho
\right)
\left(
\Pi_{\beta \nu} + \frac{1}{\sqrt{-\gammaws}} \gammaws_{\beta \gamma^\prime}\epsilon^{\gamma^\prime\delta^\prime} K_{\nu\sigma} \partial_{\delta^\prime} x^\sigma
\right)\,.
\end{align}
From the final line, we have the equation of motion for $\Pi_{\alpha \mu}$,
\be
H^{\mu\nu} \rD_\alpha \tilde x_\nu
= - \frac{1}{\sqrt{-\gammaws}} \gammaws_{\alpha \gamma}\epsilon^{\gamma\delta} H^{\mu \nu} K_{\nu \rho}  \partial_\delta x^\rho
+ H^{\mu \nu} B_{\nu\rho} \partial_\alpha x^\rho \,,
\label{onshell}
\ee
which determines $D-n-\brn$ of the combinations $\rD_\alpha \tilde x_\mu = \partial_\alpha \tilde x_\mu - \tilde\cA_{\alpha \mu}$ in terms of the physical coordinates $x^\mu$.
The remaining $n+\brn$ dual coordinates appear via the ``Lagrange multipliers''~$\beta_{\alpha i}$ and $\bar\beta_{\alpha \bri}$, which give as their equations of motion the chirality/anti-chirality constraints~\cite{Morand:2017fnv}:
\be
X_\mu^i \left(  \sqrt{-\gammaws} \gammaws^{\alpha \beta} \partial_\beta x^\mu + \epsilon^{\alpha \beta} \partial_\beta x^\mu \right) = 0 \,,\qquad
\bar X_\mu^{\bri} \left(  \sqrt{-\gammaws} \gammaws^{\alpha \beta} \partial_\beta x^\mu - \epsilon^{\alpha \beta} \partial_\beta x^\mu \right) = 0 \,,
\label{chanch}
\ee
for the directions picked out by the zero vectors $X_\mu^i$ and $\brX_\mu^{\bri}$ of the degenerate matrix $H^{\mu\nu}$.
If we eliminate $\Pi_{\alpha \mu}$ from the action \eqref{nnaction} via the equation of motion \eqref{onshell} we arrive at the analogue of the SNC action for an $(n,\brn)$ geometry, $S_{\text{DWS}}\stackrel{\text{eliminate $\Pi$}}{\xrightarrow{\hspace*{1.5cm}}} S_{(n,\brn)}$, which is obviously given by \eqref{nnaction} with the final line set to zero on imposing \eqref{onshell}, that is: \cite{Morand:2017fnv}
\begin{align}
\label{actionundoubled}
S_{(n,\brn)}
&= \dis{\frac{T}{2} \int d^2\sigma\,}- \sqrt{-\gammaws} \gammaws^{\alpha \beta} K_{\mu\nu} \partial_\alpha x^\mu \partial_\beta x^\nu
+ \epsilon^{\alpha \beta} {B}_{\mu\nu} \partial_\alpha x^\mu \partial_\beta x^\nu
\\\nonumber
&{}\qquad\qquad\qquad
+ \beta_{\alpha i} X_{\mu}^i ( - \sqrt{-\gammaws} \gammaws^{\alpha \beta} \partial_\beta x^\mu - \epsilon^{\alpha \beta} \partial_\beta x^\mu)
+ \bar \beta_{\alpha \bri} \bar X_\mu^{\bri} (- \sqrt{-\gammaws} \gammaws^{\alpha \beta} \partial_\beta x^\mu + \epsilon^{\alpha \beta} \partial_\beta x^\mu)
\,.
\end{align}

\subsubsection*{Generalised dilaton coupling}
For completeness, we may also comment on the coupling to the generalised dilaton, which is via the natural Fradkin-Tseytlin term,
\be
S_{\text{FT}} = \frac{1}{4\pi} \int d^2\sigma \sqrt{-h} R[h] d \,.
\ee
When we integrate out the components $\Pi_{\alpha \mu}$ in the path integral, we generate the usual 1-loop dilaton shift, $d \rightarrow d - \frac{1}{4} \log \det^\prime (H^{\mu\nu})$, where we take a determinant after projecting to the $D-n-\brn$ non-degenerate directions of $H^{\mu\nu}$. This cancels with the primed determinant in \eqref{nonriegendil} and leaves the usual Fradkin-Tseytlin term involving the scalar $\phi$ appearing in \eqref{nonriegendil}.

For the discussion of isometries which will we come to below, if we focus purely on the classical worldsheet, we only see the coupling to the generalised metric and hence only the Killing equations for the generalised metric~\eqref{Killing} are required, and then the full set of the infinite-dimensional symmetries~\eqref{KillingSOL} will appear.

\subsubsection*{Milne-shift invariance for the undoubled string action~\eqref{actionundoubled}}

Recall that the choice of spacetime parametrisation is not determined uniquely, but can be changed via the Milne-shift transformations~\eqref{MS}.
The  Gaussian integral of $\Pi_{\alpha\mu}$ is not a Milne-shift invariant procedure:   the  on-shell relation~\eqref{onshell} is not Milne-shift  invariant, neither is  the last line of the action~\eqref{nnaction}.
However, it is on-shell invariant up to the chirality/anti-chirality  relations \eqref{chanch} which are the equations of motion of $\beta_{\alpha i}$ and $\bar\beta_{\alpha \bri}$.
This means that starting with the reduced action \eqref{actionundoubled} the variation of the $K_{\mu\nu}$ and $B_{\mu\nu}$ couplings therein  vanishes on using these equations of motion, i.e. after imposing \emph{all} the  equations of motion of $\tilde{\cA}_{\alpha \mu}$ the Milne-shift  invariance is restored on-shell.
Alternatively, since the variation of $K_{\mu\nu}$ and $B_{\mu\nu}$ produces terms proportional to the equations of motions of $\beta_{\alpha i}$ and $\bar\beta_{\alpha \bri}$, this variation can be cancelled off-shell if in addition to the Milne variations of $K_{\mu\nu}$ and $B_{\mu\nu}$~\eqref{MS}, we let the Lagrange multipliers transform as
\begin{equation}
\label{compensate}
\begin{split}
\beta_{\alpha i}\rightarrow\beta_{\alpha i}
&+\half
\big(\partial_{\alpha}x^{\mu}-\frac{1}{\sqrt{-h}}h_{\alpha\beta}\epsilon^{\beta\gamma}\partial_{\gamma}x^{\mu}\big)
\\
&{}\qquad
\times
\left[
2(KH)_{\mu}{}^{\rho}V_{\rho i}+2V_{\rho[i}Y^{\rho}_{j]}X^{j}_{\mu}-V_{\rho i}H^{\rho\sigma}(V_{\sigma j}X^{j}_{\mu}+\brV_{\sigma \brj}\brX^{\brj}_{\mu})\right]\,,
\\
\brbeta_{\alpha\bri}\rightarrow \brbeta_{\alpha\bri}
&+\half
\big(\partial_{\alpha}x^{\mu}+\frac{1}{\sqrt{-h}}h_{\alpha\beta}\epsilon^{\beta\gamma}\partial_{\gamma}x^{\mu}\big)
\\
&{}\qquad
\times
\left[
2(KH)_{\mu}{}^{\rho}\brV_{\rho\bri}+2\brV_{\rho[\bri}\brY^{\rho}_{\brj]}\brX^{\brj}_{\mu}-\brV_{\rho\bri}H^{\rho\sigma}(V_{\sigma j}X^{j}_{\mu}+\brV_{\sigma \brj}\brX^{\brj}_{\mu})\right]
\,.
\end{split}
\end{equation}
This is \emph{not} merely the transformation induced by the use of the parametrisation in the definition of $\beta_{\alpha i}$ and $\bar\beta_{\alpha \bri}$ in terms, but can be viewed as a `compensating' transformation rule for the Lagrange multipliers.\footnote{$\tilde{\cA}_{\alpha\mu}$ must transform within \eqref{defP}, \eqref{defbetas} in order to produce \eqref{compensate}.}
Like other Milne transformations of the component fields~\eqref{MS}, the linear terms in the parameters, $V_{\mu i},\brV_{\nu\bri}$, correspond to the infinitesimal Milne-shift, $\deltaM$, and the above is the finite transformation generated by the exponentiation, $e^{\deltaM}$, which terminates at the quadratic  order.
To summarise, while the doubled sigma model of the generalised metric~\eqref{doubleWS} is trivially  invariant under the Milne transformations, the reduced undoubled action~\eqref{actionundoubled} which is free of the dual coordinates, $\tx_{\mu}$, is so provided \eqref{compensate} is taken.

\subsection{Sigma model in flat non-Riemannian geometry}

Let us now specialise to the flat $(n , \brn)$ background with coordinates $x^\mu = (x^a, x^i, \bar x^{\bri})$ and the  generalised metric \eqref{mcH0}.
In this case, after integrating out the dual coordinates $\tilde x_a$, the sigma model reads:
\be
\begin{split}
S_{\text{Flat} \,(n,\brn)} & = \frac{T}{2} \int d^2\sigma\,- \sqrt{-\gammaws} \gammaws^{\alpha \beta} \eta_{ab} \partial_\alpha x^a \partial_\beta x^b
\\ & \qquad\qquad\qquad
+ \beta_{\alpha i} ( - \sqrt{-\gammaws} \gammaws^{\alpha \beta}  - \epsilon^{\alpha \beta}) \partial_\beta x^i
+ \bar \beta_{\alpha \bri}  (- \sqrt{-\gammaws} \gammaws^{\alpha \beta}  + \epsilon^{\alpha \beta}  )\partial_\beta \brx^{\bri}
\,.
\end{split}
\label{genGO}
\ee
In conformal gauge, $\sqrt{-h} h^{00} = -1 = - \sqrt{-h} h^{11}$, $\sqrt{-h} h^{01}=0$, $\epsilon^{01}=1$, we can define\footnote{The other combinations, ${\beta_{0i} - \beta_{1i}}$, ${\bar\beta_{0\bri} + \bar\beta_{1\bri}}$, are decoupled and can be  gauged away through BRST quantisation~\cite{Park:2020ixf}.}
$\beta_i \equiv \beta_{0i} + \beta_{1i}$, $\bar\beta_{\bri} \equiv \bar\beta_{0\bri} - \bar\beta_{1\bri}$ and with $\partial_\pm = \partial_0 \pm \partial_1$ the action becomes\footnote{Without the compensating transformations \eqref{compensate}, the effect of a Milne-shift is to generate a term of the form $-\frac{T}{2} \int d^2\sigma \partial_- x^i \partial_+ \brx^{\bri} C_{i \bri}$. For $n=\brn =1$ this corresponds to a ``chemical potential'' for winding strings discussed in e.g. \cite{Ko:2015rha}, taken to be constant in \cite{Gomis:2000bd}.}
\be
\begin{split}
S_{\text{Flat} \,(n,\brn)} & = \frac{T}{2} \int d^2\sigma\, \eta_{ab} \partial_- x^a \partial_+ x^b + \beta_{i} \partial_- x^i + \bar\beta_{\bri} \partial_+ \bar x^{\bri}
\,,
\end{split}
\label{genGOconfgauge}
\ee
and so exactly generalises the Gomis-Ooguri action \eqref{eq:standard-gomis-ooguri-action}, and therefore admits $n +\bar n$ copies of the infinite-dimensional symmetry \eqref{eq:standard-gomis-ooguri-chiral-reparam} described in the Introduction. Now we demonstrate that  this is induced by the supertranslational  isometries of the generalised metric \eqref{mcH0} that we derived in the previous section.

To show this, we first discuss the transformation properties of the  Lagrange multiplier fields, $\beta,\bar \beta$.
Recalling their definitions~\eqref{defP}, \eqref{defbetas}, we directly obtain from the transformations \eqref{diffeo} that
\be
\begin{split}
\delta_{\Lambda}P_{\alpha\mu}&=-\partial_{\mu}\xi^{\nu}P_{\alpha\nu}-(\Lie_{\xi}B_{\mu\nu}+2\partial_{[\mu}\lambda_{\nu]})\partial_{\alpha}x^{\nu}\,,\\
\delta_\Lambda \beta_{\alpha i}
& = \Lie_\xi Y^\mu_i P_{\alpha \mu} - Y^\mu_i ( \Lie_\xi B_{\mu\nu} + 2 \partial_{[\mu} \lambda_{\nu]} ) \partial_\alpha x^\nu
\,,\\
\delta_\Lambda \bar \beta_{\alpha \bri}
 & = -\Lie_\xi \bar Y^\mu_{\bri} P_{\alpha \mu} + \bar Y^\mu_{\bri} ( \Lie_\xi B_{\mu\nu} + 2 \partial_{[\mu} \lambda_{\nu]} ) \partial_\alpha x^\nu \,,
\end{split}
\label{betatransfs}
\ee
where
we see the \textit{induced} transformations (via the ordinary Lie derivative $\Lie_\xi$) of the background fields, $Y^\mu_i, \bar Y^\mu_{\bri},B_{\mu\nu}$, though \textit{a priori} ${\delta_{\Lambda}Y_{i}^{\mu}=\xi^{\lambda}\partial_{\lambda}Y_{i}^{\mu}},
{\delta_{\Lambda}\brY_{\bri}^{\mu}=\xi^{\lambda}\partial_{\lambda}\brY_{\bri}^{\mu}}, {\delta_{\Lambda}B_{\mu\nu}=\xi^{\lambda}\partial_{\lambda}B_{\mu\nu}}.$
In a Riemannian setting, the Lie derivatives  would vanish when $(\lambda_\mu, \xi^\nu)$ correspond to an isometry.
However, as discussed in section \ref{isometries}, the transformations of $Y^\mu_i$, $\bY^\mu_{\bri}$ and $B_{\mu\nu}$ will only vanish up to Milne-shift and ${\GL}(n) \times {\GL}(\brn)$ rotations, reflecting the local Lorentz ambiguities inherent in the $(n,\brn)$ parametrisation.
Hence we have the non-trivial transformations of $\beta_{\alpha i}$ and $\bar \beta_{\alpha \bri}$ as in  \eqref{betatransfs}.

For the supertranslational Killing vectors  \eqref{KillingSOL}, with the on-shell  value of the dual coordinates from \eqref{onshell},   $\rD_{\alpha}\tx_{a} \equiv  - \tfrac{1}{\sqrt{-\gammaws}} \gammaws_{\alpha \gamma}\epsilon^{\gamma\beta} \partial_\beta x_a$, the general expressions \eqref{betatransfs} give
\be
\begin{split}
\delta_\Lambda \beta_{\alpha i} & = -
 \partial_i \zeta^a \big( \partial_\alpha x_a- \tfrac{1}{\sqrt{-\gammaws}} \gammaws_{\alpha \gamma}\epsilon^{\gamma\beta} \partial_\beta x_a \big)  - \partial_i \zeta^j \beta_{\alpha j}
  - 2 \partial_{[i} \rho_{j]} \partial_\alpha x^j \,,\\
 \delta_\Lambda \bar \beta_{\alpha\bri} & =
 -\bar\partial_{\bri} \bar\zeta^a \big(  \partial_\alpha x_a+\tfrac{1}{\sqrt{-\gammaws}} \gammaws_{\alpha \gamma}\epsilon^{\gamma\beta} \partial_\beta x_a  \big) - \bar \partial_{\bri} \bar\zeta^{\brj} \brbeta_{\alpha \brj}
 +2 \bar \partial_{[\bri} \bar \rho_{\brj]} \partial_\alpha \bar x^{\brj} \,.
\end{split}
\label{betaflattransfs}
\ee
One can then check that the (undoubled) action \eqref{genGO} is invariant under the transformations \eqref{betaflattransfs} along with the  supertranslational Killing vector shifts $\delta_\Lambda x^\mu = \xi^\mu$ of the coordinates~\eqref{KillingSOL}, up to a total derivative
\begin{align}
\delta_\Lambda S_{\text{Flat} \,(n,\brn)}
&= - {T} \int d^2 \sigma \epsilon^{\alpha \beta} \partial_\alpha \left( \rho_j \partial_\beta x^j + \bar\rho_{\brj} \partial_\beta \bar x^{\brj}  + (\zeta_a - \bar\zeta_a) \partial_\beta x^a  \right)
\\
&= - T \int d^2 \sigma \epsilon^{\alpha \beta} \partial_\alpha ( \lambda_\mu \partial_\beta x^\mu )
 \,.
\end{align}
In particular, when $n=\brn=1$, writing the coordinates as $x^\mu = (x^a, y,\bar y)$ and using the conformal gauge components, $\beta \equiv \beta_0 + \bar \beta_1$ and $\bar\beta\equiv \bar\beta_0 - \bar\beta_1$, the expressions \eqref{betaflattransfs} become
\be
\ba{ll}
\delta_\Lambda \beta  =
- \frac{\partial  \zeta(y)}{\partial y} \beta - 2 \frac{\partial \zeta^a(y)}{\partial y}  \partial_+ x_a \,,\qquad&\qquad
 \delta_\Lambda \bar \beta  =
 - \frac{\partial\bar \zeta(\bar y)}{\partial \bar y}  \bar \beta - 2 \frac{\partial \bar\zeta^a(\bar y)}{\partial\bar y} \partial_- x_a \,.
\ea
\ee
Indeed, this is exactly the transformation~\eqref{eq:standard-gomis-ooguri-chiral-reparam} we discussed in the Introduction, in full agreement with    \cite{Batlle:2016iel, Bergshoeff:2019pij}.

\subsection{String charges and algebra\label{SECcharges}}

\subsubsection*{Noether Charges}
We evaluate the Noether current~\eqref{stringJ} in the $(n,\brn)$ parametrisation.
Let $\Lambda^A = ( \lambda_\mu, \xi^\nu)$.
Using the equations of motion \eqref{onshell} and \eqref{chanch} we have explicitly
\be
\begin{split}
\cJ_\Lambda^\alpha & =
T \epsilon^{\alpha \beta} \lambda_\mu \partial_\beta x^\mu+ T \xi^\mu \Big( - \sqrt{-\gammaws} \gammaws^{\alpha \beta} K_{\mu\nu} \partial_\beta x^\nu + \epsilon^{\alpha \beta} B_{\mu\nu} \partial_\beta x^\nu \Big)
\\ & \quad
+  T \xi^\mu \Big[\tfrac{1}{2} X_{\mu}^{i} (- \sqrt{-\gammaws} \gammaws^{\alpha \beta} + \epsilon^{\alpha \beta} ) \beta_{\beta i}
+ \tfrac{1}{2} \bar X_{\mu}^{\bri} (-\sqrt{-\gammaws} \gammaws^{\alpha \beta}-\epsilon^{\alpha \beta} ) \bar\beta_{\beta \bri}
\Big]
\,.
\end{split}
\label{Jonshell}
\ee
For the flat space  Killing vector solution~\eqref{KillingSOL}, this Noether current~\eqref{Jonshell} becomes
\be
\begin{split}
\frac{1}{T} \mathcal{J}_\Lambda^\alpha & =
\epsilon^{\alpha \beta}
\left(
\partial_\beta \varphi +
\rho_i( x^k) \partial_\beta x^i
+
\bar \rho_{\bri}( \bar x^{\bar k}) \partial_\beta  \bar x^{\bri}
\right)  - \omega_{ab} x^b \sqrt{-\gammaws} \gammaws^{\alpha \beta}  \partial_\beta x^a
\\ & \qquad
+ \zeta_a(x^k)\left( - \sqrt{-\gammaws} \gammaws^{\alpha \beta} +  \epsilon^{\alpha \beta}  \right) \partial_\beta x^a
+ \bar \zeta_a(\bar x^{\bar k}) \left(   -  \sqrt{-\gammaws} \gammaws^{\alpha \beta}-\epsilon^{\alpha \beta} \right) \partial_\beta x^a
\\& \qquad
+ \frac{1}{2} \zeta^i (x^k) (- \sqrt{-\gammaws} \gammaws^{\alpha \beta} + \epsilon^{\alpha \beta} ) \beta_{\beta i}
+ \frac{1}{2} \bar \zeta^{\bri}(\bar x^{\bar k}) (-\sqrt{-\gammaws} \gammaws^{\alpha \beta} -\epsilon^{\alpha \beta} ) \bar\beta_{\beta \bri}
\,.
\end{split}
\label{Jonshellflat}
\ee
Note that
through the projections with $- \sqrt{-\gammaws} \gammaws^{\alpha \beta}\pm \epsilon^{\alpha \beta}$,
the Noether currents associated to $\zeta_{a},\zeta^{i}$ and $\brzeta_{a},\brzeta^{\bri}$ are chiral and anti-chiral, respectively, on the worldsheet.

We define a conserved charge, $\mathcal{Q}_\Lambda \equiv
\oint d\sigma \cJ_\Lambda^0\,$.
In conformal gauge, this reads
\be
\begin{split}
\mathcal{Q}_\Lambda & = \oint \Big[
\omega_{ab} p^{[a} x^{b]}
+ \zeta^a (x^k) ( p_a + T x^{\prime}_a )
+ \bar \zeta^a(\bar x^{\bar k}) ( p_a - Tx^\prime_a )
+  \zeta^i (x^k) p_i
+  \bar\zeta^{\bri}( \bar x^{\bar k}) \bar p_{\bri}\\ & \qquad \qquad
+
T \left(
\varphi^\prime +
 \rho_i( x^k)  x^{\prime i}
+
\bar \rho_{\bri}( \bar x^{\bar k})   \bar x^{\prime \bri}\right)
\Big] \,,
\end{split}
\label{TheQ}
\ee
where we substituted for the momenta conjugate to $x^a$, $x^i$ and $\bar x^{\bri}$,
\be
p_a \equiv T \dot{x}_a \equiv  T \eta_{ab} \dot{x}^b \,,\quad
p_i \equiv \frac{T}{2} \beta_i\,,\quad
\bar p_{\bri} \equiv \frac{T}{2} \bar\beta_{\bri}\,.
\ee
Henceforth we set $T=1$ for convenience.

\subsubsection*{Algebra}

From the charge \eqref{TheQ}, we can extract the generators of the isometry algebra: for the usual Lorentz rotations,
\be
M_{ab}=\dis{\int\rd\sigma~}p_{a}x_{b}-p_{b}x_{a}\,,
\ee
and for the supertranslations,\footnote{Imposing \eqref{constantc}, both $P_{i}^{\vl}$ and $\brP_{\bri}^{\vbrl}$ may be subject to further constraints which can be easily implemented.}
\begin{alignat}{1}
P_{a}^{\vl}=\dis{\int\rd\sigma~}(x^{1})^{l_{1}}(x^{2})^{l_{2}}
\cdots (x^{n})^{l_{n}}(p_{a}+x^{\prime b}\eta_{ba})\,,\quad&\quad
\brP_{a}^{\vbrl}=\dis{\int\rd\sigma~}(\brx^{1})^{\brl_{1}}(\brx^{2})^{\brl_{2}}
\cdots (\brx^{\brn})^{\brl_{\brn}}(p_{a}-x^{\prime b}\eta_{ba})\,,
\nonumber\\
P_{i}^{\vl}=\dis{\int\rd\sigma~}(x^{1})^{l_{1}}(x^{2})^{l_{2}}
\cdots (x^{n})^{l_{n}}p_{i}\,,\quad&\quad
\brP_{\bri}^{\vbrl}=\dis{\int\rd\sigma~}(\brx^{1})^{\brl_{1}}(\brx^{2})^{\brl_{2}}
\cdots (\brx^{\brn})^{\brl_{\brn}}\brp_{\bri}\,,
\\\nonumber
R^{i}_{\vl}=\dis{\int\rd\sigma~}(x^{1})^{l_{1}}(x^{2})^{l_{2}}
\cdots (x^{n})^{l_{n}} x^{\prime i}\,,\quad&\quad
\brR^{\bri}_{\vbrl}=\dis{\int\rd\sigma~}(\brx^{1})^{\brl_{1}}(\brx^{2})^{\brl_{2}}
\cdots (\brx^{\brn})^{\brl_{\brn}}\brx^{\prime\bri}\,,
\end{alignat}
where we set $n$-dimensional and $\brn$-dimensional vector notations,
\be
\ba{ll}
\vl=(l_{1},l_{2},\cdots,l_{n})\,,\qquad&\qquad
\vbrl=(\brl_{1},\brl_{2},\cdots,\brl_{\brn})\,,
\ea
\ee
and let the components, $l_{i}$'s, $\brl_{\bri}$'s,  be non-negative integers. If any of them is negative, the corresponding generator vanishes.
To write the commutators it is  convenient to introduce unit vectors, $\hati,\hatbri,\hatj,\hatbrj,\hatk,\hatbrk$,  such that for example,
\be
\vl-\hati=(l_{1},\cdots,l_{i}-1,\cdots,l_{n})\,,\quad
\vbrm-\hatbrk=(\brm_{1},\cdots,\brm_{\brk}-1,\cdots,\brm_{\brn})\,.
\ee
Using this notation, the nontrivial commutation relations are:
\be
\frac{1}{i}\big[M_{ab}\,,\,M_{cd}\big]=
\eta_{cb}M_{ad}-\eta_{ca}M_{bd}+\eta_{db}M_{ca}-\eta_{da}M_{cb}\,,
\ee
\begin{alignat}{2}
\frac{1}{i}\left[M_{ab}\,,\,P^{\vl}_{c}\,\right]&=\eta_{cb}P^{\vl}_{a}-\eta_{ca}P^{\vl}_{b}\,,\quad&\quad
\frac{1}{i}\left[M_{ab}\,,\,\brP^{\vbrl}_{c}\,\right]&=\eta_{cb}\brP^{\vbrl}_{a}-\eta_{ca}\brP^{\vbrl}_{b}\,,
\nonumber\\
\frac{1}{i}\left[P^{\vl}_{a}\,,\,P^{\vm}_{b}\,\right]&=\sum_{k=1}^{n}\eta_{ab}(m_{k}-l_{k})R^{k}_{\vl+\vm-\hatk}\,,\quad&\quad
\frac{1}{i}\left[\brP^{\vbrl}_{a}\,,\,\brP^{\vbrm}_{b}\,\right]&=\sum_{\brk=1}^{\brn}\eta_{ab}(\brl_{\brk}-\brm_{\brk})\brR^{\brk}_{\vbrl+\vbrm-\hatbrk}\,,
\nonumber\\
\frac{1}{i}\left[P^{\vl}_{a}\,,\,P^{\vm}_{i}\,\right]&=l_{i}P_{a}^{\vl+\vm-\hati}\,,\quad&\quad
\frac{1}{i}\left[\brP^{\vbrl}_{a}\,,\,\brP^{\vbrm}_{\bri}\,\right]&=\brl_{\bri}\brP_{a}^{\vbrl+\vbrm-\hatbri}\,,
\\\nonumber
\frac{1}{i}\left[P^{\vl}_{i}\,,\,P^{\vm}_{j}\,\right]&=-m_{i}P_{j}^{\vl+\vm-\hati}+
l_{j}P^{\vl+\vm-\hatj}_{i}\,,\quad&\quad
\frac{1}{i}\left[\brP^{\vbrl}_{\bri}\,,\,\brP^{\vbrm}_{\brj}\,\right]&=-\brm_{\bri}\brP_{\brj}^{\vbrl+\vbrm-\hatbri}+
\brl_{\brj}\brP^{\vbrl+\vbrm-\hatbrj}_{\bri}\,,
\end{alignat}
\be
\begin{aligned}
\frac{1}{i}\left[P^{\vl}_{i}\,,\,R^{j}_{\vm}\,\right]=-m_{i}R^{j}_{\vl+\vm-\hati}+{\sum_{k=1}^{n}\,}\delta_{i}^{~j}m_{k}
R^{k}_{\vl+\vm-\hatk}\,,\\
\frac{1}{i}\left[\brP^{\vbrl}_{\bri}\,,\,\brR^{\brj}_{\vbrm}\,\right]=-\brm_{\bri}
\brR^{\brj}_{\vbrl+\vbrm-\hatbri}+{\sum_{\brk=1}^{\brn}\,}\delta_{\bri}^{~\brj}\brm_{\brk}
\brR^{\brk}_{\vbrl+\vbrm-\hatbrk}\,.
\end{aligned}
\label{COMPR}
\ee
All other commutators are trivial. In particular,  the chiral (unbarred) and the anti-chiral (barred) supertranslational  generators commute: suppressing indices,
\be
\big[\,P\,,\,\brP\,\big]=0\,,\quad\big[\,P\,,\,\brR\,\big]=0\,,\quad
\big[\,R\,,\,\brP\,\big]=0\,,\quad \big[\,R\,,\,\brR\,\big]=0\,.
\ee
It is worthwhile to note
\be
{\sum_{k=1}^{n}\,}l_{k}R^{k}_{\vl-\hatk}=0\,,\quad\qquad
{\sum_{\brk=1}^{\brn}\,}\brl_{\brk}\brR^{\brk}_{\vbrl-\hatbrk}=0\,,
\ee
since these correspond to the integrals of total derivatives. We then have identities like
\be
{\sum_{k=1}^{n}\,}m_{k}
R^{k}_{\vl+\vm-\hatk}={\sum_{k=1}^{n}\,}\half(m_{k}-l_{k})
R^{k}_{\vl+\vm-\hatk}\,,
\ee
and can verify the consistency between \eqref{COM2} and \eqref{COMPR}. Furthermore, with $\vec{0}=(0,0,\cdots,0)$, we note $P^{\vec{0}}_{a}$ and $\brP^{\vec{0}}_{a}$ coincide,\footnote{Similarly  $R^{i}_{\vec{0}}$ and $\brR^{\bri}_{\vec{0}}$ vanish trivially. Otherwise they would have formed an $(n+\brn)$-dimensional ideal. }
\be
P^{\vec{0}}_{a}=\brP^{\vec{0}}_{a}
=\dis{\int\rd\sigma~}p_{a}\,,
\ee
and  commute with all other  supertranslational generators,
\be
\left[P^{\vec{0}}_{a}\,,\,P\,\right]=0\,,\quad
\left[P^{\vec{0}}_{a}\,,\,\brP\,\right]=0\,,\quad
\left[P^{\vec{0}}_{a}\,,\,R\,\right]=0\,,\quad
\left[P^{\vec{0}}_{a}\,,\,\brR\,\right]=0\,.
\ee
Thus, they form  the usual quadratic Casimir operators for  `mass squared':\footnote{In \eqref{signm}, the minus sign in front of $m^{2}$ can be neglected if $\eta^{ab}$ is not  mostly plus Minkowskian.}
\be
\left<P^{\vec{0}}_{a}P^{\vec{0}}_{b}\eta^{ab}\right>=\left<
\brP^{\vec{0}}_{a}\brP^{\vec{0}}_{b}\eta^{ab}\right>\equiv -m^{2}\,.
\label{signm}
\ee
Namely, only the Riemannian directions are involved.

Some comments are in order. \vspace{-8pt}
\begin{itemize}
\item[--] The presence of the infinitely many  conserved Noether charges  means the integrability of the `free'  string action on the flat non-Riemannian background~\eqref{genGO}. Chiral or anti-chiral strings satisfy ${x^{i}(\tau,\sigma)=x^{i}(0,\tau+\sigma)}$ or ${\brx^{\bri}(\tau,\sigma)=\brx^{\bri}(0,\tau-\sigma)}$, which imply   they   are fixed in space and preserve their  whole  shapes.
\vspace{-8pt}
\item[--]If $n=1$ or $\brn=1$, $R^{i}_{\vl}$ or $\brR^{\bri}_{\vbrl}$  vanishes trivially and we recover the algebra of \cite{Batlle:2016iel}. In the above, we have neglected any possible winding numbers.
Restoring these will result in additional extensions, and it would be interesting to compare the result with~\cite{Batlle:2016iel} and \cite{Bergshoeff:2019pij}.\vspace{-8pt}
\item[--]Instead of the string~\eqref{doubleWS}, if the generalised metric is coupled to a point particle~\cite{Ko:2016dxa}, it is straightforward to derive the  worldline Noether charge for the non-Riemannian isometries,
\be
Q_{\Lambda}(x,p)=\left[w^{a}{}_{b}x^{b}+\zeta^{a}(x^{k})+\brzeta^{a}(\brx^{\brk})
\right]p_{a}+\zeta^{i}(x^{k})p_{i}+\brzeta^{\bri}(\brx^{\brk})\brp_{\bri}\,.
\label{NCphase}
\ee
Compared with the string case of \eqref{TheQ}, this expression lacks the information of the dual (tilde) directions,  realising only  the (untilde) commutation relations~\eqref{COM1}  while missing the tilde part~\eqref{COM2}. This is of course consistent with the  intuition that  the doubled geometry is intrinsically stringy rather than point particle-like.

\end{itemize}

\section{Discussion\label{CONSection}}
The generalised metric of double field theory (DFT) provides a unified description of Riemannian and non-Riemannian geometries, and in this paper we showed how this description can be applied to the notion of Killing symmetries.
In particular, we showed that flat non-Riemannian spacetime admits an infinite-dimensional algebra of supertranslational isometries.
We also showed how these symmetries extend to the supersymmetric theory, which gives rise to Killing spinors corresponding to arbitrary chiral and anti-chiral reparametrisations.

\paragraph{Curved geometries}
The double geometry approach advocated in the above can also be used to study the properties of general curved non-Riemannian backgrounds.
The infinite-dimensional isometries that we  found for flat backgrounds will not persist for a general background,
but the appropriate DFT realisation of the Killing equations in Equations~\eqref{Killing} and~\eqref{Kd} still applies.
As a first example, in  Appendix~\ref{AppendixCurved} we consider the case of the non-Riemannian geometry originally obtained  in \cite{Lee:2013hma} (which is related by simultaneous  time- and space-like  T-duality to the fundamental string supergravity solution).
In that case, the Killing equations do not lead to an infinite-dimensional isometry group, but instead give the global isometries $\mathbf{O}(8)\times\mathbf{ISO}(1,1)$.

\paragraph{Non-relativistic strings}
For the non-Riemannian parametrisation corresponding to torsional Newton--Cartan (TNC) or stringy Newton--Cartan (SNC) strings, the bosonic infinite-dimensional symmetries have previously been obtained from the string worldsheet action~\cite{Batlle:2016iel,Bergshoeff:2019pij}.
In this work, we show that they can in fact be attributed to isometries of the non-relativistic background geometry.
We have not taken winding modes into account in the worldsheet realisation of these symmetries, which may add extensions to their algebra.
It would be interesting to fix these extensions and their associated transformations in particular in the context of non-relativistic strings, since they were previously argued~\cite{Bergshoeff:2018yvt,Bergshoeff:2019pij} to correspond to a particular torsion (or foliation) constraint for SNC geometry.
This constraint is relevant when comparing known SNC string beta functions~\cite{Gomis:2019zyu,Yan:2019xsf} to the recent computation of the DFT effective equations of motion in terms of the non-Riemannian parametrisations~\cite{Cho:2019ofr,Gallegos:2020egk}.
Likewise, building on our computation in Appendix~\ref{AppendixCurved}, it would be interesting to study the breaking of the infinite-dimensional isometries in general curved non-relativistic string backgrounds in more detail.
Finally, a doubled perspective is likely to be useful also for the non-relativistic open strings that were recently considered in~\cite{Gomis:2020fui,Gomis:2020izd}.

\paragraph{Boundary charges}
For studying the properties of more general non-Riemannian backgrounds, another potentially useful tool is the DFT construction of boundary charges~\cite{Blair:2015eba,Park:2015bza,Naseer:2015fba}.
Analogously to (for example)
the ADM charge in standard general relativity, this construction associates a conserved boundary charge to a global generalised Killing vector field.
For this, it is useful to rewrite the generalised Killing equations \eqref{Killing} and \eqref{Kd} in terms of the (torsionless) covariant derivative of DFT as~\cite{Park:2015bza}\footnote{The vielbein  Killing equations~\eqref{KVV} can be re-expressed  in terms of covariant derivatives too~\cite{Angus:2018mep}, see Eq.(3.3) therein.}
\be
\hcL_{\Lambda}\cH_{AB}
=8\brP_{(A}{}^{C}P_{B)}{}^{D}\na_{[C}\Lambda_{D]}=0\,,\qquad
\hcL_{\Lambda}d
=-\half\na_{A}\Lambda^{A}=0\,.
\label{Killing2}
\ee
For a given Killing vector fulfilling these two conditions, the contractions with the DFT Einstein curvature, or (on-shell) equivalently with the DFT energy-momentum tensor, are conserved~\cite{Angus:2018mep},
\be
\na_{A}(G^{AB}\Lambda_{B})=0=\na_{A}(T^{AB}\Lambda_{B})\quad\Longleftrightarrow\quad\partial_{A}(e^{-2d}G^{AB}\Lambda_{B})=0=\partial_{A}(e^{-2d}T^{AB}\Lambda_{B})\,.
\label{contractedJ}
\ee
These target space conserved currents are comparable with the worldsheet Noether current~\eqref{stringJ}.
The generalised boundary charge associated to a doubled  Killing vector $\Lambda^A$ is obtained by integrating~\cite{Blair:2015eba,Park:2015bza}
\be
\mathcal{Q}^{AB} =
e^{-2 d} \left[ 4 (\brP^{C[A}P^{B]D} -P^{C[A}\brP^{B]D})\nabla_C \Lambda_D  - 2 N^{[A} \Lambda^{B]} \right] \,,
\label{ADMDFT}
\ee
where the DFT boundary vector \cite{Berman:2011kg} is $N^A = -\partial_B \cH^{AB} + 4 \cH^{AB} \partial_B d$.
In practice, one wants to integrate this over a codimension-2 hypersurface within the physical spacetime.
Solving the section condition by letting ${\tpartial^{\mu}\equiv0}$,   we only need  the components $\mathcal{Q}^{\mu\nu}$.
In terms of the differential  ``toolkit'' for the undoubled non-Riemannian geometry that was introduced in~\cite{Cho:2019ofr},  we can easily find the corresponding expression in the general $(n,\bar n)$ case, with $\Lambda^{M}=(\lambda_{\mu},\xi^{\nu})$,
\be
\mathcal{Q}^{\mu\nu} = e^{-2d} \left[
- 2 \hat{\mathfrak{D}}^{[\mu} \xi^{\nu]} - (\lambda_\rho-B_{\rho\sigma}\xi^{\sigma}) \hat{\mathbb{H}}^{\mu\nu\rho}
+ (\partial_\rho H^{\rho [\mu} - 4 H^{\rho[\mu} \partial_\rho d ) \xi^{\nu]}
\right]\,,
\ee
where $\hat{\mathfrak{D}}^\mu$ is a covariant derivative and $\hat{\mathbb{H}}^{\mu\nu\rho}$ is a Milne-shift invariant $H$-flux.
(All upper-indexed, see section 4.3 of \cite{Cho:2019ofr} for their precise definitions.)
It would be interesting to see what the resulting charges are for curved non-Riemannian geometries.

\paragraph{Relation to asymptotic symmetries}
The infinite-dimensional Killing isometries that we obtained  are strongly reminiscent of the BMS algebra associated to the reparametrisation symmetries of null infinity in asymptotically flat spacetimes~\cite{Bondi:1962px,Sachs:1962wk,Sachs:1962zza}.
This is where we borrowed the term `supertranslations'.
We emphasise that the supertranslational symmetries we have studied do not appear only asymptotically, but are  the genuine  isometries of the whole   non-Riemannian spacetime we considered.
Though  we have considered the non-Riemannian geometries without any reference to an embedding in a higher-dimensional space, utilising the DFT Kaluza-Klein ansatz, one can obtain, e.g.  the Carrollian parametrisation as a null hypersurface in a higher-dimensional Riemannian DFT geometry~\cite{Morand:2017fnv}.
Likewise, it would be worthwhile to explore the appropriate notion of boundary conditions and associated asymptotic symmetries, building on the existing toolkit for computing boundary charges that was discussed above.

\paragraph{Supersymmetry and M-theory}
We have used the known supersymmetric formulation of DFT to obtain the generalised Killing spinor equations for the non-Riemannian parametrisation  with $n=\brn=1$ and $D=10$ that is related to non-relativistic strings.
DFT also provides us with a maximally supersymmetric effective low-energy action~\cite{Jeon:2012hp}, and it would be interesting to work out the  supersymmetric action for a general non-Riemannian parametrisation  including the Ramond-Ramond sector.
On the worldsheet side, a good starting point would be the spacetime or worldsheet supersymmetric doubled sigma models considered in this context in~\cite{Park:2016sbw, Blair:2019qwi}.
Finally, the extension of the analysis of generalised isometries to M-theoretic non-Riemannian backgrounds can be considered following \cite{Berman:2019izh}.

\paragraph{Other directions} In the conventional flat  Minkowskian spacetime, `particles' are identified as  the  irreducible representations of the Poincar\'{e} group.
It would be of interest to generalise Wigner's classification to the non-Riemannian geometries considered above.
In particular, the notion of particle mass is valid as long as there is a Minkowskian subspace, see~\eqref{signm}.

An especially intriguing application of the non-Riemannian geometries may be, as   an alternative to string compactifications,  to assume the internal space to be non-Riemannian while keeping the external  four-dimensional spacetime Riemannian as usual~\cite{Morand:2017fnv,Cho:2018alk,Park:2020ixf}.
Our result then  appears to indicate   that,  when the internal non-Riemannian space is flat, there will be supertranslational symmetries  in  effective field theories around the total background.
These are local and hence should be taken as gauge symmetries.
That is to say, physical states in this background (at the quantum level)  should be  supertranslational singlets.
Specifically, in view of our Killing vector solution \eqref{KillingSOL},  they should have no dependence on the internal non-Riemannian space.
This could imply that non-Riemannian isometries  provide  a natural scheme for the dimensional reduction(s) from the critical ten (or 26) to the phenomenological four dimensions.

\section*{Acknowledgements}
We wish to thank Eric Bergshoeff, Emanuel Malek, Kevin Morand, Niels Obers and Shigeki Sugimoto for useful discussions.
We are grateful to the authors of \cite{Gallegos:2020egk} for sharing a draft of their paper with us and for further discussions.
We also acknowledge the organizers of the conference \href{https://workshops.aei.mpg.de/geometryduality/}{\textit{Geometry and Duality, Potsdam,
December 2-6, 2019}} where our collaboration was initiated.

CB is supported by the FWO-Vlaanderen through a Senior Postdoctoral Fellowship and through the project G006119N, and is also supported by the Vrije Universiteit Brussel through the Strategic Research Program ``High-Energy Physics''.
GO is supported by the project “Towards a deeper understanding of black holes with non-relativistic holography” of the Independent Research Fund Denmark (grant number DFF-6108-00340) and the Villum Foundation Experiment project 00023086.
JHP  is  supported by Basic Science Research Program through the National Research Foundation of Korea (NRF)   through  the Grants,  NRF-2016R1D1A1B01015196 and NRF-2020R1A6A1A03047877.

\appendix

\section{Solving the flat non-Riemannian Killing equations
\label{derivKilling}}

Here we derive the most general  Killing vector solution~\eqref{KillingSOL} for the generalised metric \eqref{mcH0}.
With $\cH_{AB}$ constant, it is convenient to return to the generalised Killing equations in the form \eqref{genKillingGeneral}, which  become explicitly:
\beqa
&&\partial_{\rho}\xi^{\mu}\cH^{\rho\nu}+\partial_{\rho}\xi^{\nu}\cH^{\mu\rho}=0\,,\label{K1}\\
&&\partial_{\mu}\xi^{\rho}\cH_{\rho}{}^{\nu}+\partial_{\mu}\lambda_{\rho}\cH^{\rho\nu}
-\partial_{\rho}\lambda_{\mu}\cH^{\rho\nu}
-\partial_{\rho}\xi^{\nu}\cH_{\mu}{}^{\rho}=0\,,\label{K2}\\
&&\partial_{\mu}\xi^{\rho}\cH_{\rho\nu}+\partial_{\mu}\lambda_{\rho}\cH^{\rho}{}_{\nu}
-\partial_{\rho}\lambda_{\mu}\cH^{\rho}{}_{\nu}+
\partial_{\nu}\xi^{\rho}\cH_{\mu\rho}+
\partial_{\nu}\lambda_{\rho}\cH_{\mu}{}^{\rho}-
\partial_{\rho}\lambda_{\nu}\cH_{\mu}{}^{\rho}=0\,.\label{K3}
\eeqa
The generalised Killing equations \eqref{K1} to \eqref{K3} now further decompose into ${3\times 3=9}$ sets of equations  in view of \eqref{flatcoords}, depending on the free Greek indices, $\mu,\nu$, being of $a$, $i$, or $\bri$ type.
For the constant $(n,\brn)$ background, the first set of the Killing  equations~\eqref{K1} we are going to solve becomes
\be
\eta^{c(\mu}\partial_{c}\xi^{\nu)}=0\,.
\label{start}
\ee
When $\{\mu,\nu\}=\{a,b\}$, we recover the  Poincaré symmetry: from
\be
\partial_{a}\xi_{b}+\partial_{b}\xi_{a}=0\,,
\label{abxi}
\ee
we have as usual,
\begin{equation}
  \begin{split}
    &\partial_{c}\partial_{a}\xi_{b}=-\partial_{c}\partial_{b}\xi_{a}=\partial_{b}\partial_{a}\xi_{c}=
    -\partial_{a}\partial_{c}\xi_{b}=0\quad
    \\
    &{}\qquad\qquad\Longrightarrow
    \xi_{a}(x^{c},x^{k},\brx^{\brk})=w_{ab}(x^{k},\brx^{\brk})x^{b}+v_{a}(x^{k},\brx^{\brk})\,,
  \end{split}
  \label{zetaa}
\end{equation}
where $w_{ab}=-w_{ba}$ is skew-symmetric, and the  $a,b,c,d$ indices are freely raised or lowered by $\eta^{ab}$ or $\eta_{cd}$. On the other hand, if $\{\mu,\nu\}=\{a,i\}$ or $\{a,\bri\}$,  we have\footnote{
In our notation, $\{\mu,\nu\}$ is unordered while $(\mu,\nu)$ is ordered: for example   ${\{\mu,\nu\}=\{a,i\}}$ means ${\mu=a},\,{\nu=i}$ or  ${\mu=i},\,{\nu=a}$, while ${(\mu,\nu)=(a,i)}$ denotes  ${\mu=a},\,{\nu=i}$.}
\be
\ba{lllll}
\partial_{c}\xi^{i}=0\,,&~~
\partial_{c}\brxi^{\bri}=0\quad&\Longrightarrow&\quad
\xi^{i}=\zeta^{i}(x^{k},\brx^{\brk})\,,&~~
\brxi^{\bri}=\brzeta^{\bri}(x^{k},\brx^{\brk})\,.
\ea
\label{zetaibri}
\ee
Other cases of $\{\mu,\nu\}$ being $\{i,j\}$, $\{i,\brj\}$, or $\{\bri,\brj\}$ are trivial implying no constraint.  That is to say, the most general solutions to the first equation~\eqref{K1} with constant $\cH^{\mu\nu}=H^{\mu\nu}$ are given by the superrotations and supertranslations~\eqref{zetaa}, \eqref{zetaibri},
\be
\cL_{\xi}H^{\mu\nu}=0
\quad\Longleftrightarrow\quad\xi^{\mu}(x^{c},x^{k},\brx^{\brk})=\left[w^{a}{}_{b}(x^{k},\brx^{\brk})x^{b}+v^{a}(x^{k},\brx^{\brk})\,,\,\zeta^{i}(x^{k},\brx^{\brk})\,,\,\brzeta^{\bri}(x^{k},\brx^{\brk})\right].
\label{superSOL}
\ee
After acquiring this, we turn to the second set of the Killing equations~\eqref{K2} which  reads
\be
\partial_{\mu}\zeta^{i}\delta_{i}{}^{\nu}-
\partial_{\mu}\brzeta^{\bri}\delta_{\bri}{}^{\nu}+
(\partial_{\mu}\lambda_{c}-\partial_{c}\lambda_{\mu})\eta^{c\nu}
-\partial_{i}\xi^{\nu}\delta_{\mu}{}^{i}+\brpartial_{\bri}\xi^{\nu}\delta_{\mu}{}^{\bri}=0\,.
\label{K2p}
\ee
If $(\mu,\nu)=(a,b)$, we have $
\partial_{[a}\lambda_{b]}=0$ and hence with an arbitrary   function~$\tvarphi(x^{c},x^{k},\brx^{\brk})$,
\be
\lambda_{a}=\partial_{a}\tvarphi(x^{c},x^{k},\brx^{\brk})\,.
\ee
The cases of $(\mu,\nu)=(a,i)$, $(a,\bri)$, $(i,j)$, or $(\bri,\brj)$ are trivial. If $(\mu,\nu)=(i,\bri)$ or $(\bri,i)$ we  note the   `chiral' property,
\be
\ba{lllll}
\partial_{i}\brzeta^{\bri}(x^{k},\brx^{\brk})=0\,,&~~
\brpartial_{\bri}\zeta^{i}(x^{k},\brx^{\brk})=0\quad&\Longrightarrow&\quad
\brxi^{\bri}=\brzeta^{i}(\brx^{\brk})\,,&~~
\xi^{i}=\zeta^{i}(x^{k})\,.
\ea
\label{zetaibriholo}
\ee
If $(\mu,\nu)=(i,a)$, \eqref{K2p} gives
\be
\partial_{a}\lambda_{i}(x^{c},x^{k},\brx^{\brk})=\partial_{i}\!\left[\partial_{a}\tvarphi(x^{c},x^{k},\brx^{\brk})-v_{a}(x^{k},\brx^{\brk})-w_{ab}(x^{k},\brx^{\brk})x^{b}\right]\,,
\label{K2pia}
\ee
of which the integrability condition of $\partial_{[a}\partial_{b]}\lambda_{i}=0$ implies $\partial_{i}w_{ab}=0$.
 Consequently,  the last term in \eqref{K2pia} drops out to give
\be
\partial_{a}\lambda_{i}(x^{c},x^{k},\brx^{\brk})=\partial_{a}\partial_{i}\!\left[\tvarphi(x^{c},x^{k},\brx^{\brk})-x^{c}v_{c}(x^{k},\brx^{\brk})\right]\,,
\ee
and thus,
\be
\lambda_{i}(x^{c},x^{k},\brx^{\brk})=\partial_{i}\!\left[\tvarphi(x^{c},x^{k},\brx^{\brk})-x^{c}v_{c}(x^{k},\brx^{\brk})\right]+\alpha_{i}(x^{k},\brx^{\brk})\,,
\label{lambdai}
\ee
where $\alpha_{i}(x^{k},\brx^{\brk})$ is an arbitrary function of $x^{k}$ and $\brx^{\brk}$.  Similarly for $(\mu,\nu)=(\bri,a)$, we get $
{\brpartial_{\bri}w_{ab}=0}$ and
\be
\brlambda_{\bri}(x^{c},x^{k},\brx^{\brk})=\brpartial_{\bri}\!\left[\tvarphi(x^{c},x^{k},\brx^{\brk})+x^{c}v_{c}(x^{k},\brx^{\brk})\right]+\bralpha_{\bri}(x^{k},\brx^{\brk})\,.
\label{lambdabri}
\ee
In particular, the skew-symmetric parameter, $w_{ab}$, must be strictly constant. That is to say, there is no  `superrotation' (of the Riemannian space) left after imposing the second equation~\eqref{K2}.

The last set of the Killing equations~\eqref{K3} now  reduces to
\be
\partial_{(\mu}\xi^{c}\eta_{\nu) c}
+\delta^{i}{}_{(\mu}\partial_{\nu)}\lambda_{i}
-\delta^{\bri}{}_{(\mu}\partial_{\nu)}\brlambda_{\bri}
-\partial_{i}\lambda_{(\mu}\delta^{i}{}_{\nu)}+
\brpartial_{\bri}\lambda_{(\mu}\delta^{\bri}{}_{\nu)}=0\,.
\ee
For $\{\mu,\nu\}=\{a,b\}$, we only recover \eqref{abxi} and nothing new.
Likewise, the cases of  $\{\mu,\nu\}=\{a,i\}$ or $\{a,\bri\}$ are automatically fulfilled  by \eqref{zetaa}, \eqref{lambdai}, and \eqref{lambdabri}.
The cases of $\{\mu,\nu\}=\{i,j\}$ and $\{\bri,\brj\}$ are trivial.
The remaining final case of $\{\mu,\nu\}=\{i,\bri\}$  gives
\be
\partial_{i}\brlambda_{\bri}-\brpartial_{\bri}\lambda_{i}=2x^{c}\partial_{i}\brpartial_{\bri}v_{c}(x^{k},\brx^{\brk})+\partial_{i}\bralpha_{\bri}(x^{k},\brx^{\brk})-
\brpartial_{\bri}\alpha_{i}(x^{k},\brx^{\brk})=0\,.
\ee
This   implies
\be
\ba{ll}
\partial_{i}\brpartial_{\bri}v_{c}(x^{k},\brx^{\brk})=0\,,\qquad&\qquad
\brpartial_{\bri}\alpha_{i}(x^{k},\brx^{\brk})-\partial_{i}\bralpha_{\bri}(x^{k},\brx^{\brk})=0\,,
\ea
\ee
which are generically solved by
\be
\ba{l}
v_{a}(x^{k},\brx^{\brk})=\zeta_{a}(x^{k})+\brzeta_{a}(\brx^{\brk})\,,\\
\alpha_{i}(x^{k},\brx^{\brk})=\partial_{i}\sigma(x^{k},\brx^{\brk})+\rho_{i}(x^{k})\,,\\
\bralpha_{\bri}(x^{k},\brx^{\brk})=\brpartial_{\bri}\sigma(x^{k},\brx^{\brk})+\brrho_{\bri}(\brx^{\brk})\,.
\ea
\ee
Here   $\zeta_{a}(x^{k}),\rho_{i}(x^{k})$ are `chiral' and  $\brzeta_{a}(\brx^{\brk}),\brrho_{\bri}(\brx^{\brk})$ are  `anti-chiral', while $\sigma(x^{k},\brx^{\brk})$ is a scalar depending on both $x^{k}$ and $\brx^{\brk}$ but not on $x^{a}$. As for the final step, we perform a field redefinition,
\be
\ba{lll}
\tvarphi(x^{c},x^{k},\brx^{\brk})&\longrightarrow&
\varphi(x^{c},x^{k},\brx^{\brk}):=\tvarphi(x^{c},x^{k},\brx^{\brk})-x^{a}\zeta_{a}(x^{k})
+x^{a}\brzeta_{a}(\brx^{\brk})+\sigma(x^{k},\brx^{\brk})\,,
\ea
\label{frr}
\ee
which removes $\sigma$ by absorption and simplifies our general solution. This completes our derivation of \eqref{KillingSOL}.\\

\section{Solving for a curved non-Riemannian geometry \label{AppendixCurved}}

The first known non-Riemannian geometry of DFT is given by the following generalised metric~\cite{Lee:2013hma}
\be
\cH_{MN} =\left(\ba{cccc}
0&0&  \sigma_{3}&0\\
0&\delta^{ab}&0&0\\
\sigma_{3} &0 & f\sigma_{1}&0\\
0&0&0&\delta_{cd}\ea\right)\,,\qquad f= 1+\frac{Q}{r^{6}}\,,\qquad r^{2} = \sum_{a=1}^{8} (x^{a})^{2}\,,
\label{nonRgeo}
\ee
while the generalised dilaton, $d$, is constant.
This can be obtained by starting with the usual fundamental string supergravity solution~\cite{Dabholkar:1990yf}, with metric $ds^2 = f^{-1} (-dt^2+dz^2) + \delta_{ab} dx^a dx^b$, $B$-field $B_{tz} = f^{-1}$, and dilaton $e^{-2\phi} =f$, and formally T-dualising on both $t$ and $z$.

The corresponding   Killing equations  then consist of
\beqa
&&\partial_{\rho}\xi^{\mu}\cH^{\rho\nu}+\partial_{\rho}\xi^{\nu}\cH^{\mu\rho}=0\,,\label{K1c}\\
&&\partial_{\mu}\xi^{\rho}\cH_{\rho}{}^{\nu}+\partial_{\mu}\lambda_{\rho}\cH^{\rho\nu}
-\partial_{\rho}\lambda_{\mu}\cH^{\rho\nu}
-\partial_{\rho}\xi^{\nu}\cH_{\mu}{}^{\rho}=0\,,\label{K2c}\\
&&
\xi^{\rho}\partial_{\rho}\cH_{\mu\nu}+\partial_{\mu}\xi^{\rho}\cH_{\rho\nu}+\partial_{\mu}\lambda_{\rho}\cH^{\rho}{}_{\nu}
-\partial_{\rho}\lambda_{\mu}\cH^{\rho}{}_{\nu}+
\partial_{\nu}\xi^{\rho}\cH_{\mu\rho}+
\partial_{\nu}\lambda_{\rho}\cH_{\mu}{}^{\rho}-
\partial_{\rho}\lambda_{\nu}\cH_{\mu}{}^{\rho}=0\,,\label{K3c}
\eeqa
and  for the dilaton, $d$,
\be
\partial_{\mu}\xi^{\mu}=0\,.
\label{Kdc}
\ee
Compared to the flat cases of \eqref{K1}, \eqref{K2}, \eqref{K3}, and \eqref{volume}, the only difference is the transport term, $\xi^{\rho}\partial_{\rho}\cH_{\mu\nu}$,  in \eqref{K3c}, which is nontrivial due to  the harmonic function positioned in $\cH_{\mu\nu}$ of  \eqref{nonRgeo}. Each of \eqref{K1c}, \eqref{K2c}, and \eqref{K3c} decomposes into ${3\times 3=9}$ sets of equations, depending on the  Greek indices, $\mu,\nu$, being $+$,  $-$, or $a=1,2,\cdots,8$. We naturally put
\be
\ba{ll}
\xi^{\mu}= \left(\xi^{+}\,,\,\xi^{-}\,,\,\xi^{a}\right)\,,\quad&\quad
\lambda_{\nu}= \left(\lambda_{+}\,,\,\lambda_{-}\,,\,\lambda_{b}\right)\,,
\ea
\ee
and solve for these variables \textit{a priori} as functions of $x^{+},x^{-},x^{c}$.  Luckily, since \eqref{K1c} and \eqref{K2c} are identical to \eqref{K1} and \eqref{K2}, all the previous analyses spanning from \eqref{start} to \eqref{lambdabri} for the flat cases are readily applicable:   the most general solution to \eqref{K1c}, \eqref{K2c} are given by
\be
\ba{lll}
\xi^{+}=\zeta^{+}(x^{+})\,,\qquad&\qquad
\xi^{-}=\zeta^{-}(x^{-})\,,\qquad&\qquad
\xi^{a}=w^{a}{}_{b}x^{b}+v^{a}(x^{+},x^{-})\,,
\ea
\label{ready1}
\ee
and
\be
\ba{l}
\lambda_{+}=\partial_{+}\big[\tvarphi(x^{c},x^{+},x^{-})-x^{c}v_{c}(x^{+},x^{-})\big]+\alpha_{+}(x^{+},x^{-})\,,\\
\lambda_{-}=\partial_{-}\big[\tvarphi(x^{c},x^{+},x^{-})+x^{c}v_{c}(x^{+},x^{-})\big]+\alpha_{-}(x^{+},x^{-})\,,\\
\lambda_{a}=\partial_{a}\tvarphi(x^{c},x^{+},x^{-})\,.
\ea
\label{ready2}
\ee
We only need to take care of  \eqref{K3c} and \eqref{Kdc} henceforth. With \eqref{ready1} and \eqref{ready2},  the Killing equation~\eqref{K3c} is  nontrivial only for the case of $\{\mu,\nu\}=\{+,-\}$, to give
\be
\xi^{c}\partial_{c}f+f\partial_{+}\zeta^{+}+f\partial_{-}\zeta^{-}-4x^{c}\partial_{+}\partial_{-}v_{c}
+2\partial_{-}\alpha_{+}-2\partial_{+}\alpha_{-}=0\,,
\ee
which, from $\partial_{c}f=-6Qx_{c} r^{-8}$,   further reduces to
\be
\begin{split}
  &4\partial_{[-}\alpha_{+]}(x^{+},x^{-})
  +f(r)\left[\partial_{+}\zeta^{+}(x^{+})+\partial_{-}\zeta^{-}(x^{-})\right]
  \\
  &{}\qquad\qquad
  -4x^{c}\partial_{+}\partial_{-}v_{c}(x^{+},x^{-})
  -6Qr^{-8}x^{c}v_{c}(x^{+},x^{-})=0\,.
\end{split}
\label{reduced}
\ee
There are four terms here which have  distinct dependence on  the eight-dimensional coordinates, $x^{c}$, when $Q\neq 0$.  Therefore, each of them should vanish separately, or
\be
\ba{lll}
\partial_{[-}\alpha_{+]}(x^{+},x^{-})=0\,,\qquad&\qquad
\partial_{+}\zeta^{+}(x^{+})+\partial_{-}\zeta^{-}(x^{-})=0\,,\qquad&\qquad
v_{a}(x^{+},x^{-})=0\,.
\ea
\label{eachseparate}
\ee
For example,  one may take  a partial derivative, $\partial_{a}$, of \eqref{reduced}, and obtain
\be
\begin{split}
  &\partial_{+}\partial_{-}v_{a}(x^{+},x^{-})
  \\
  &{}\qquad
  +\frac{3}{2}Qr^{-8}\left[x_{a}\partial_{+}\zeta^{+}(x^{+})+x_{a}\partial_{-}\zeta^{-}(x^{-})+v_{a}(x^{+},x^{-})-8r^{-2}x_{a}x^{c}v_{c}(x^{+},x^{-})\right]=0\,.
\end{split}
\ee
Considering the large $r$ limit, we get $\partial_{+}\partial_{-}v_{a}(x^{+},x^{-})=0$, and hence
\be
x_{a}\partial_{+}\zeta^{+}(x^{+})+x_{a}\partial_{-}\zeta^{-}(x^{-})+v_{a}(x^{+},x^{-})-8r^{-2}x_{a}x^{c}v_{c}(x^{+},x^{-})=0\,.
\ee
Again taking  a partial derivative, $\partial_{b}$, of this and considering large $r$ limit, we note $\partial_{+}\zeta^{+}+\partial_{-}\zeta^{-}=0$ and subsequently,
\be
r^{2}v_{a}(x^{+},x^{-})-8x_{a}x^{c}v_{c}(x^{+},x^{-})=0\,.
\ee
Finally, hitting this with $\partial_{b}\partial^{a}$, we get $v_{b}=0$. The first relation in \eqref{eachseparate} is solved by $\alpha_{\pm}=\partial_{\pm}\sigma(x^{+},x^{-})$. In conclusion, for $Q\neq 0$, with the field redefinition, $\varphi\equiv\tvarphi+\sigma$ like \eqref{frr}, the most general solution to the full set of the Killing equations is
\be
\xi^{a}=w^{a}{}_{b}x^{b}\,,\qquad
\xi^{+}=\omega x^{+}+c^{+}\,,\qquad
\xi^{-}=-\omega x^{-}+c^{-}\,,\qquad
\lambda_{\mu}=\partial_{\mu}\varphi\,,
\label{KillingSOLc}
\ee
which corresponds to  the direct product of $\so(8)$ and
the two-dimensional Poincaré symmetry of  the light-cone.\\

\noindent In contrast to the flat case \eqref{KillingSOL}, there is no supertranslation.
Furthermore, the Killing equations for the DFT metric implies the dilatonic Killing equation: the general solution to the former is now already divergenceless, $\partial_{\mu}\xi^{\mu}=0$, while in the flat case the dilatonic Killing equation imposes a separate constraint, see~\eqref{volume}.
The lack of infinite-dimensional isometries may well  imply that the corresponding interacting string action~\eqref{actionundoubled} on this background is not integrable.

\bibliographystyle{JHEP}
\bibliography{minimal}

\end{document}